%% file: main.tex
\documentclass[journal]{IEEEtran}
\IEEEoverridecommandlockouts
\usepackage[noadjust]{cite}
\usepackage[hidelinks,pageanchor=false,hypertexnames=false]{hyperref}
\usepackage{booktabs}
\usepackage{amsmath}
\usepackage{amssymb}
\usepackage{mathtools}
\usepackage{bbm}
\usepackage{bm}
\usepackage{amsfonts}
\usepackage{accents}
\usepackage{comment}
\usepackage{amsthm}

\newtheorem{lemma}{Lemma}
\newtheorem{proposition}{Proposition}

\usepackage[capitalize]{cleveref}
\crefname{appsec}{Appendix}{Appendices}

\DeclareMathOperator*{\argmax}{arg\,max}
\DeclareMathOperator*{\E}{\mathbb{E}}

\DeclareMathOperator{\Tr}{Tr}

\newcommand*\df{\mathop{}\!\mathrm{d}}
\usepackage{pgfplots}
\pgfplotsset{compat=1.18}
\usetikzlibrary{fadings}
\usetikzlibrary{patterns}
\usetikzlibrary{plotmarks}
\usepgfplotslibrary{groupplots}
\pgfdeclarelayer{background}
\pgfdeclarelayer{foreground}
\pgfsetlayers{background,main,foreground}
\usepgfplotslibrary{fillbetween}
\usetikzlibrary{automata,positioning}
\usetikzlibrary{external}
\usepackage{xparse}
\NewDocumentCommand{\includetikzimage}{ r<> r<> }{%
  \tikzsetnextfilename{#1}%
  \begin{tikzpicture}[#2]%
    \input{figs/#1.tex}%
  \end{tikzpicture}%
}

\usepackage[caption=false]{subfig}

\renewcommand{\Cref}[1]{\cref{#1}}
\usepackage{algorithm}
\usepackage{algpseudocode}
\usepackage{balance}

\begin{document}
\title{Data Sourcing Random Access using\\ Semantic Queries for Massive IoT Scenarios}

\author{Anders~E.~Kal{\o}r,~\IEEEmembership{Member,~IEEE},
	Petar~Popovski,~\IEEEmembership{Fellow,~IEEE}, and
	Kaibin Huang,~\IEEEmembership{Fellow,~IEEE}%
        \thanks{This paper was presented in part at the 21st International Symposium on Modeling and Optimization in Mobile, Ad Hoc, and Wireless Networks (WiOpt), 2023~\cite{kalorwiopt23}.}%
        \thanks{The work of A.~E.~Kal{\o}r and P.~Popovski was supported in part by the SNS JU project 6G-GOALS under the EU's Horizon program Grant Agreement No 101139232 and in part by the Velux Foundation, Denmark, through the Villum Investigator Grant WATER, nr. 37793. The work of A.~E.~Kal{\o}r was also supported in part by the Independent Research Fund Denmark (IRFD) under Grant 1056-00006B, and in part by the Japan Science and Technology Agency (JST) ASPIRE program (grant no. JPMJAP2326).The work of K.~Huang was supported in part by the Research Grants Council of the Hong Kong Special Administrative Region, China under a fellowship award (HKU RFS2122-7S04), NSFC/RGC CRS (CRS\_HKU702/24), the Areas of Excellence scheme grant (AoE/E-601/22-R), Collaborative Research Fund (C1009-22G), and the Grants 17212423 \& 17304925, and in part by the Shenzhen-Hong Kong-Macau Technology Research Programme (Type C) (SGDX20230821091559018).  }%
        \thanks{A.~E.~Kal{\o}r is with the Department of Information and Computer Science, Keio University, Yokohama 223-8522, Japan, and also with the Department of Electronic Systems, Aalborg University, 9220 Aalborg, Denmark (e-mail: aek@keio.jp).}%
        \thanks{P.~Popovski is with the Department of Electronic Systems, Aalborg University, 9220 Aalborg, Denmark (e-mail: petarp@es.aau.dk).}%
        \thanks{K.~Huang is with the Department of Electrical and Electronic Engineering, The University of Hong Kong, Hong Kong (e-mail: huangkb@eee.hku.hk).}%
}%

\maketitle

\begin{abstract}
  Efficiently retrieving relevant data from massive Internet of Things (IoT) networks is essential for downstream tasks such as machine learning. This paper addresses this challenge by proposing a novel data sourcing protocol that combines semantic queries and random access. The key idea is that the destination node broadcasts a semantic query describing the desired information, and the sensors that have data matching the query then respond by transmitting their observations over a shared random access channel, for example to perform joint inference at the destination. However, this approach introduces a tradeoff between maximizing the retrieval of relevant data and minimizing data loss due to collisions on the shared channel. We analyze this tradeoff under a tractable Gaussian mixture model and optimize the semantic matching threshold to maximize the number of relevant retrieved observations. The protocol and the analysis are then extended to handle a more realistic neural network-based model for complex sensing. Under both models, experimental results in classification scenarios demonstrate that the proposed protocol is superior to traditional random access, and achieves a near-optimal balance between inference accuracy and the probability of missed detection, highlighting its effectiveness for semantic query-based data sourcing in massive IoT networks.
\end{abstract}

\begin{IEEEkeywords}
Semantic communication, edge inference, distributed sensing, Internet of things, massive random access.
\end{IEEEkeywords}

\section{Introduction}
Wireless Internet of Things (IoT) devices are becoming increasingly advanced and equipped with a multitude of sensors, allowing them to monitor complex signals, such as images, video, and audio/speech~\cite{chiang2016fogiot}. Enhanced by advanced processing in the cloud or on an edge server, usually relying on machine learning (ML) and artificial intelligence (AI), IoT devices can be used for a wide range of applications, such as autonomous control, surveillance, detection of accidents, fault detection in critical infrastructure (e.g., smart grids and water networks), and many others~\cite{massiveiot6g,popovski2022timing}. However, collecting the massive, continuous stream of data produced by the sensors over a capacity-constrained wireless channel while considering the strict power constraints of the IoT devices poses a significant challenge. Furthermore, only a small fraction of the monitored data may be relevant for applications at the destination node~\cite{kountouris21semanticemp,6ggoals}. For example, a wildlife monitoring application may only be interested in tracking a particular animal at a given time. Similarly, data relevant for an industrial monitoring application may depend on data observed by other sensors. For instance, a sudden temperature spike in a machine component might only be relevant if accompanied by abnormal vibration readings from other sensors.

\begin{figure*}
  \centering
  \includetikzimage<protocol><font=\footnotesize,every node/.style={inner sep=0,outer sep=0}>
  \caption{The proposed protocol for data sourcing random access. (1) The edge server constructs a semantic query, $\bm{q}$ based on the query example $\bm{x}_q$ and broadcasts it in the first frame slot; (2) Each device computes a matching score $\bar{\chi}(\bm{x}_m,\bm{q})$ characterizing how well its observation $\bm{x}_m$ matches the query; (3) The devices whose matching score exceeds a threshold transmit their observations in the remaining slots using a random access protocol, and the edge server executes an application, such as inference, based on the received observations.}\label{fig:protocol}
\end{figure*}

Conventional IoT protocols rely on device-initiated \emph{push} communication, such as periodic reporting through random access, and are often agnostic to data relevance and significance, which can lead to inefficient use of resources~\cite{uysal22semantic}. Furthermore, even if the devices implement more intelligent transmission policies that depend on their data content, the devices have limited information about the global state of the network and the significance of the data within the context of the application, preventing them from making optimal transmission decisions. Alternatively, communication can be \emph{pull}-based, where the destination node actively requests data from specific devices based on what is needed by the application. The simplest form of pull-based communication is scheduling, where individual devices are granted dedicated, contention-free resources to transmit their observations. However, with generic pull-based communication, the destination and scheduler are unaware of the actual data available at the devices, creating the risk of requesting data that ultimately turns out to be irrelevant.

To minimize irrelevant transmissions, this paper considers \emph{query}-driven communication, which is a specific instance of pull-based communication. Rather than scheduling individual devices, the destination node transmits a query that specifies which kind of information the application seeks to obtain. Thus, our proposed scheme can be viewed as a novel semantic access barring method, where access decisions are guided by data relevance to a specific query rather than by conventional probabilistic or device-centric rules. Despite introducing a small overhead due to the query transmission (in the order of 8--128 bytes), this brings the advantage of collecting only relevant data, and is an attractive scheme when only a small fraction of the sensors produce relevant data, and when the sensors with relevant data are hard to predict. Since the devices with relevant information form a random subset, the query in the downlink is followed by random access transmission of the sensing data. This coupling of semantic filtering and contention-based access creates the new and fundamental challenge of balancing the probability of \emph{missed detection} (MD) and \emph{false alarm} (FA) at the semantic level while simultaneously minimizing the risk of data loss caused by random access collisions at the physical level. In this paper, we analyze this tradeoff and optimize the protocol parameters to maximize the performance.

Motivated by recent advancements in semantic communication, focusing on enabling the destination to make decisions rather than reconstructing the transmitted message exactly as in conventional communication~\cite{kountouris21semanticemp}, we propose a semantic query construction. Specifically, our aim is to design a query that enable the devices to determine whether their observation is semantically relevant given the query and meaningful within the context of the application. As an illustration, consider an IoT network of cameras deployed for wildlife monitoring. To track, for instance, a reported wild bear, a conventional push-based communication where cameras transmit images periodically, or, e.g., when there is movement, suffers from low capture probability and potential staleness of information. Conversely, in our proposed semantic query-driven, pull-based approach, the edge server first broadcasts a ``bear'' query to the devices. This query could, for instance, be a quantized feature vector constructed using a semantic ``query-by-example'' approach~\cite{rasiwasia07querybyexample} as illustrated in \cref{fig:protocol}, where the query is constructed from an image of a bear. Using the received query, the IoT cameras compute a \emph{matching score} that characterizes how well their captured image matches the query. They then transmit their captured image over the random access channel only if the matching score is sufficiently high. This results in higher efficiency and timeliness. Note that specific query class (``bear'') need not be directly observable by neither the edge server nor the devices, since it is implicitly described by the semantic query. The end objective at the server after receiving the captured images could, for instance, be to track the location of the bear, or to identify or classify the specific bear. We will consider both cases in this paper.

Previous works have explored various ways to incorporate data relevance in transmission policies. Recently, metrics like age of information (AoI) and its variants~\cite{yates2021aoisurvey,ayan2019age,maatouk2020aoii,chiariotti2022qaoi} have been used in both push and pull communication to ensure relevance of the collected data by taking into account its timeliness at the destination node. However, timeliness may not be sufficient to quantify data relevance for all types of applications, such as the ones mentioned above. Furthermore, these works often rely on simple data models at the devices, or require detailed, continuously updated state information at the destination, which is challenging in practice, especially when communication is infrequent. Another recent direction is semantic and goal-oriented communication, introduced earlier, which aims to optimize data transmission by considering the meaning and purpose of the communicated information at the destination~\cite{kountouris21semanticemp,6ggoals,lan2021semantic}. Notable examples of semantic communication include protocols for image transmission that aims to maximize the perception quality of the received image~\cite{erdemir23jscc,huang25djscc,zhou25feature}, protocols for edge inference where the goal of the destination is to perform inference on the source~\cite{lan23progressftx,xu23edge,wen24sensingai}, and an image transmission scheme that enables the destination to retrieve images from a local database that are similar to the one observed by the transmitter~\cite{jankowski21retrieval}.
Semantic and goal-oriented communication has also been applied to filter streaming IoT sensing data, where individual devices independently assess the semantic importance of their data~\cite{agheli24semfiltering}. As mentioned earlier, this approach is constrained by the lack of information available at the devices. Furthermore, compared to existing work on semantic communication, which focuses on recovering semantic information, our focus is not on encoding but on sensor selection based on semantic relevance to the query (i.e., semantic matching).

The concept of query-driven communication has been considered in, e.g., in-network query processing, where logical queries are used to filter data~\cite{kapadia2006comparative,yao2022cougar}. Other examples include the use of content-based wake-up radios to collect the top-$k$ values in a sensor network~\cite{shiraishi2020content}. Related to semantic filtering,~\cite{shiraishi24tinyairnet} proposes to occasionally transmit ML models to the devices to enable them to locally assess the relevance of their observations. This approach can be seen as a special case of our work, in which the ML models serve as (large) queries. Closest to our work is the concept of \emph{semantic data sourcing}, which was introduced in~\cite{huang2023semdas} and further analyzed in~\cite{liu25semdas}. As in this paper, the concept in~\cite{huang2023semdas,liu25semdas} relies on a semantic query that summarizes the desired information and enables each device to decide the relevance of its data. Based on the query and their current sensing data, each device computes a matching score, which is sent back to the edge server. The edge server then schedules a subset of the devices to transmit their full observations. By leveraging ML to construct the query, the authors demonstrate how semantic data sourcing can reduce the communication efficiency of numerous edge-AI use cases.

The main limitation of the protocol in~\cite{huang2023semdas,liu25semdas} is the need to collect matching scores from all devices, which is costly when the number of devices is large. This paper addresses this limitation by proposing a more scalable joint data sourcing and random access protocol where devices with a matching score exceeding a predefined threshold directly transmit their observations over a shared random access channel (see \cref{fig:protocol}). Our protocol can be realized using any random access scheme, since the design of the semantic query controls the probability of node activation in the random access process. As a specific protocol, we have focused on general irregular repetition slotted ALOHA (IRSA)~\cite{liva11irsa}, which represents a modern massive random access scheme and generalizes conventional slotted ALOHA widely used in commercially available IoT protocols~\cite{munari21aloha}. Following random access, the edge server can then process the received observations for tasks, such as machine learning inference as in~\cite{huang2023semdas,liu25semdas}. While our primary objective is joint classification (e.g., a multi-view inference scenario~\cite{su2015multiview}), we will also characterize more general retrieval quantities such as the probability of MD and FA.

Our main contributions and findings can be summarized as follows:
\emph{(i)} We propose the concept of data sourcing random access using semantic queries, which tightly integrates pull-based semantic data sourcing with contention-based IRSA random access in the uplink to overcome the scalability issues of existing schemes.
\emph{(ii)} We characterize the fundamental tradeoffs between the observation discrimination gain, query dimensionality, MD and FA by analyzing the problem under a tractable Gaussian mixture model (GMM) for sensing.
\emph{(iii)}  We propose an efficient method to optimize the matching score threshold to minimize the probability of missed detection. The numerical results show that the proposed method provides near-optimal threshold selection across several tasks.
\emph{(iv)}  We present experiments that show that the insights obtained using the GMM are, generally, consistent with results observed from a more realistic convolutional neural network (CNN)-based model applied to visual sensing data.
While our analysis and experiments are centered around visual sensing data and focus on addressing the communication bottleneck, our method can be straightforwardly applied to other settings and implemented on resource-constrained devices such as microcontrollers using, e.g., TinyML~\cite{lin2020mcunet}. However, we defer the study of computational complexity and latency to future work.

The remainder of the paper is organized as follows. \Cref{sec:system-model-problem} presents the general setup and the system model. We analyze the system under GMM sensing in \Cref{sec:semantic_matching}, and optimize the matching score in \Cref{sec:threshold}. In \cref{sec:neuralnet}, we extend the framework to a more realistic ML-based sensing scenario, in which sensors observe images and the edge server aims to perform image classification. \Cref{sec:results} presents numerical results, and the paper is concluded in \Cref{sec:conclusion}.

\textit{Notation:} Vectors are represented as column vectors and written in lowercase boldface (e.g., $\bm{x}$), while matrices are written in uppercase boldface (e.g., $\mathbf{X}$). $\Gamma(\cdot)$, $\Gamma(\cdot,\cdot)$, and $Q_{M}(\alpha,\beta)$ denote the gamma function, the upper incomplete gamma function, and the generalized Marcum Q-function of order $M$, respectively. The multivariate Gaussian probability density function (PDF) with mean $\boldsymbol{\mu}$ and covariance matrix $\boldsymbol{\Sigma}$ evaluated at $\bm{x}$ is written $\mathcal{N}(\bm{x}|\boldsymbol{\mu},\boldsymbol{\Sigma})$. $\mathbbm{1}[\cdot]$ is the indicator function, evaluating to  $1$ for a true argument and $0$ otherwise. The symbols are summarized in \cref{tab:notation}.

\begingroup
\setlength{\tabcolsep}{2pt}
\begin{table}[tb]
\centering\footnotesize
\caption{Symbols used in the paper}\label{tab:notation}\vspace{-1em}
\begin{tabular}{lp{0.67\columnwidth}}
\toprule
Symbol & Description \\\midrule
$M$ & Number of sensors \\
$\mathcal{Z}$ & Set of data classes \\
$z_q\in \mathcal{Z}$ & Class of the server's query \\
$z_m\in \mathcal{Z}$ & Class of the data observed by device $m$ \\
$\boldsymbol{x}_q\in\mathbb{R}^d$ & Feature vector for the query class \\
$\boldsymbol{x}_m\in\mathbb{R}^d$ & Feature vector for the data at device $m$ \\
$\boldsymbol{q}\in\mathbb{R}^l$ & Semantic query vector broadcast by the server \\
$\chi(\boldsymbol{x}_m, \boldsymbol{x}_q)$ & Relevance score of $\boldsymbol{x}_m$ given $\boldsymbol{x}_q$ \\
$\bar{\chi}(\boldsymbol{x}_m, \boldsymbol{q})$ & Matching score between $\boldsymbol{x}_m$ and $\boldsymbol{q}$\\
$\tau$ & Semantic matching threshold \\
$p_{\text{pos}}$ & Probability that a device observes the query class $z_q$ \\
$\mathcal{M}_{rx}\subseteq\{1,\ldots,M\}$ & Set of devices from which observations are received \\
$\bar{\boldsymbol{x}}\in\mathbb{R}^d$ & Fused feature vector at the server \\
$\boldsymbol{\mu}_z\in\mathbb{R}^d, \boldsymbol{\Sigma}\in\mathbb{R}^{d\times d}$ & Mean and covariance for class $z$ in the GMM \\
$\xi_{\boldsymbol{\Sigma}}(\boldsymbol{a}, \boldsymbol{b})$ & Squared Mahalanobis distance between $\boldsymbol{a}$ and $\boldsymbol{b}$ under covariance $\boldsymbol{\Sigma}$ \\
$f_{\text{enc}}(\cdot)$ & Feature encoder model (CNN) \\
$f_{\text{ser}}(\cdot)$ & Server's classifier model \\
$\hat{z}\in\mathcal{Z}$ & Final class estimate after inference \\
$\varepsilon_{\text{MD}}$ & Probability of missed detection\\
$\varepsilon_{\text{FA}}$ & Probability of false alarm \\
$p_{\text{err}}^{\text{dl}}$ & Query transmission error probability \\
$p_{\text{err}}^{\text{ul}}$ & Uplink transmission error probability \\
$L_{ul}$ & Number of slots in the uplink random access frame \\
$\Lambda(x)$ & Degree distribution polynomial for IRSA \\
\bottomrule
\end{tabular}
\end{table}
\endgroup

\section{System Model}\label{sec:system-model-problem}
We consider a scenario with $M$ IoT sensor devices connected through a wireless link to an edge server. Time is divided into independent recurring frames, each comprising a semantic matching phase and a random access phase, as depicted in \cref{fig:frame_model}. In the semantic matching phase, the edge server broadcasts a semantic query based on a given \emph{query example}~\cite{rasiwasia07querybyexample}, with the aim of allowing each of the IoT devices to determine whether its current sensing data are relevant for the server. The devices that conclude from the semantic matching that their sensing data are relevant transmit them over a random access channel to the edge server. While our protocol can be applied to many applications, we assume that the overall goal is to retrieve and then perform joint classification of the sensor observations that are relevant, according to a certain criterion, to the query example. We present the sensing, inference, and communication models, as well as the considered performance metrics in the sequel.

\subsection{Sensing and Inference Model}\label{sec:obs_model}
We assume that the query example at the edge server represents a \emph{query class} $z_q\in\mathcal{Z}=\{1,2,\ldots,|\mathcal{Z}|\}$ drawn uniformly at random from the set of classes $\mathcal{Z}$. Since $\mathcal{Z}$ can be very large in practice, we assume that it is known only to the edge server and not to the IoT devices. The query class $z_q$ is assumed not to be directly observable, but only partially observable through the query example. Based on the query example, the edge server extracts a \emph{query feature vector} $\bm{x}_q\sim p(\bm{x}_q|z_q),\ \bm{x}_q\in\mathbb{R}^{d}$, e.g., using a neural network, which it uses to construct the semantic query (elaborated later). Here, the distribution $p(\bm{x}_q|z_q)$ arises from the fact that $\bm{x}_q$ is constructed from a random query example representing query class $z_q$. Similarly, each IoT device, indexed by $m=1,2,\ldots,M$, produces a feature vector $\bm{x}_m\in\mathbb{R}^{d}$ representing its sensed data of some class $z_m\in\mathcal{Z}$, which is also not directly observable (i.e., only through the sensed data).
We assume that each of the $M$ devices produces a feature vector of the query class $z_q$ with probability $p_{\mathrm{pos}}$, or otherwise produces a feature vector of a different class drawn uniformly from the set of remaining classes $\mathcal{Z}\setminus z_q$. Letting $\mathsf{Z}=(z_q,z_1,\ldots,z_M)$ denote the query class and the class observed by each of the $M$ devices, the joint class distribution can be written
\begin{equation*}
  p(\mathsf{Z})=\frac{1}{|\mathcal{Z}|}\prod_{m=1}^M \left(p_{\mathrm{pos}}\right)^{\mathbbm{1}[z_m=z_q]}\left(\frac{1-p_{\mathrm{pos}}}{|\mathcal{Z}|-1}\right)^{\mathbbm{1}[z_m\neq z_q]}.
\end{equation*}
We denote the indices of the devices that observe class $z\in\mathcal{Z}$ by $\mathcal{M}_{z}$. To capture correlation between sensors observing the same class (e.g., as in multi-view sensing~\cite{su2015multiview}), we assume that the feature vectors $\mathsf{X}=(\bm{x}_q,\bm{x}_1,\ldots,\bm{x}_M)$ are drawn from the joint distribution
\begin{equation}
  p(\mathsf{X}|\mathsf{Z})=p(\bm{x}_q|z_q)\prod_{z=1}^{|\mathcal{Z}|}\int_{\kappa_z}\prod_{i\in\mathcal{M}_z}p(\bm{x}_i|\kappa_z)p(\kappa_z|z)\,\mathrm{d}\kappa_z,\label{eq:feature_dist}
\end{equation}
where $\kappa_z$ is a latent variable characterizing the correlation among the sensors observing class $z$. In words, the feature vectors observed by the devices $i\in\mathcal{M}_z$ that observe class $z$ are drawn independently from the conditional distribution $p(\mathbf{x}_i|\kappa_z)$, where $\kappa_z\sim p(\kappa_z|z)$, common for all devices observing class $z$, describes the correlation among the device feature vectors. The specific distribution $p(\kappa_z|z)$ will be elaborated in \cref{sec:gaussian_sensing_model,sec:cnn_sensing_model}. It follows that the marginal feature vector distribution $p(\mathsf{X})$ is given as $p(\mathsf{X})=\sum_{\mathsf{Z}} p(\mathsf{X}|\mathsf{Z})p(\mathsf{Z})$.

Recall that only the devices whose semantic matching scores exceed the threshold $\tau$ will transmit their features to the server. After the feature transmission, we assume that the edge server computes a fused feature vector from the received feature vectors, as
\begin{equation}
  \bar{\bm{x}} = \frac{\sum_{m\in\mathcal{M}_{\mathrm{rx}}} \chi(\bm{x}_m,\bm{x}_q)\bm{x}_m}{\sum_{m'\in\mathcal{M}_{\mathrm{rx}}} \chi(\bm{x}_{m'},\bm{x}_q)},\label{eq:feature_fusion}
\end{equation}
where $\mathcal{M}_{\mathrm{rx}}$ is the set of devices from which it has received feature vectors and $\chi(\bm{x}_m,\bm{x}_q)\ge 0$ is the weight given to feature $\bm{x}_m$ given the query $\bm{x}_q$. The weight function $\chi(\bm{x}_m,\bm{x}_q)$ is assumed to represent the relevance of the observation given the query, and depends on the specific problem. This type of feature fusion resembles an attention mechanism, and has been used previously in the literature~\cite{vaswani17attention,when2com,where2comm}.
We consider two specific sensing models as described next.

\subsubsection{GMM for Linear Classification}\label{sec:gaussian_sensing_model}
To get insights into the problem, we primarily focus on an analytically tractable Gaussian sensing model, where all feature vectors are drawn independently from Gaussian distributions conditioned on the observed classes as
\begin{equation*}
  p\left(\mathsf{X}|\mathsf{Z}\right)=\mathcal{N}(\bm{x}_q|\boldsymbol{\mu}_{z_q},\boldsymbol{\Sigma})\prod_{i=1}^{M}\mathcal{N}(\bm{x}_i|\boldsymbol{\mu}_{z_i},\boldsymbol{\Sigma}).
\end{equation*}
Note that the means (or class centroids) are determined by the class observed by each sensor, $z_i$, while the covariance matrix $\boldsymbol{\Sigma}\in\mathbb{R}^{d\times d}$ is assumed to be the same for all classes. We assume that $\boldsymbol{\Sigma}$ is diagonalized with diagonal entries $C_{1,1},C_{2,2},\ldots,C_{d,d}$ (e.g., obtained using singular value decomposition). Note that the marginal feature vector distribution follows a GMM with density
\begin{equation}\label{eq:gaussian_sensing_marginal}
  p\left(\bm{x}_i\right)=\frac{1}{|\mathcal{Z}|}\sum_{z=1}^{|\mathcal{Z}|}\mathcal{N}(\bm{x}_i|\boldsymbol{\mu}_{z},\boldsymbol{\Sigma}).
\end{equation}

As relevancy metric, we consider the exponential squared Mahalanobis distance between a feature vector $\bm{x}_m$ and the query feature vector $\bm{x}_q$, defined as
\begin{align}
  \chi(\bm{x}_m,\bm{x}_q)&=e^{-\frac{1}{d}\xi_{\boldsymbol{\Sigma}}(\bm{x}_m,\bm{x}_q)},\label{eq:gaussian_similarity_score}
\end{align}
where
\begin{align}
  \xi_{\boldsymbol{\Sigma}}(\bm{a},\bm{b})&=\left(\bm{a}-\bm{b}\right)^T\boldsymbol{\Sigma}^{-1}\left(\bm{a}-\bm{b}\right)\nonumber\\
  &
  =\sum_{n=1}^d \frac{(\bm{a}[n]-\bm{b}[n])^2}{C_{n,n}}\label{eq:squared_mahalan}
\end{align}
is the squared Mahalanobis distance between $\bm{a}$ and $\bm{b}$ under covariance matrix $\boldsymbol{\Sigma}$. Note that the factor $(1/d)$ in the exponent of \cref{eq:gaussian_similarity_score} normalizes the exponent by the variance of the Mahalanobis distance when $\bm{x}_m$ and $\bm{x}_q$ are drawn from the same class.\footnote{We define the relevancy score as in \cref{eq:gaussian_similarity_score} rather than, e.g., the Mahalanobis distance directly, since \eqref{eq:gaussian_similarity_score} is normalized between zero and one and its expected value does not grow unboundedly with the feature dimension $d$. However, we note that this relevance score is not necessarily optimal.}

Complete maximum a posteriori classification of the fused feature vector is challenging because of the correlation introduced by the weight functions. Instead, we treat the fused feature vector $\bar{\bm{x}}$ as a weighted average of $|\mathcal{M}_{\mathrm{rx}}|$ feature vectors drawn from the \emph{same} class under the assumption that the feature weights are independent of the feature vectors themselves. Under this assumption, the fused feature vector $\bar{\bm{x}}$ is drawn from a GMM equivalent to that in \cref{eq:gaussian_sensing_marginal} with class means
\begin{align*}
  \widehat{\boldsymbol{\mu}}_z&= \frac{\sum_{m\in\mathcal{M}_{\mathrm{rx}}} \chi(\bm{x}_m,\bm{x}_q)\bm{\mu}_{z}}{\sum_{m'\in\mathcal{M}_{\mathrm{rx}}} \chi(\bm{x}_{m'},\bm{x}_q)}=\bm{\mu}_{z},
\end{align*}
and covariance
\begin{align*}
  \widehat{\boldsymbol{\Sigma}}&=\sum_{m\in\mathcal{M}_{\mathrm{rx}}}\left(\frac{\chi(\bm{x}_m,\bm{x}_q)}{\sum_{m'\in\mathcal{M}_{\mathrm{rx}}} \chi(\bm{x}_{m'},\bm{x}_q)}\right)^2\boldsymbol{\Sigma}\\
  &=\left(\frac{ \sum_{m\in\mathcal{M}_{\mathrm{rx}}}\chi(\bm{x}_m,\bm{x}_q)^2}{\left(\sum_{m'\in\mathcal{M}_{\mathrm{rx}}} \chi(\bm{x}_{m'},\bm{x}_q)\right)^2}\right)\boldsymbol{\Sigma}.
\end{align*}
Note that this choice of model ensures that the variance of the individual feature entries decreases as the number of observations increases. The a posteriori distribution is then
\begin{align*}
  p_{\mathrm{pred}}(z|\bar{\bm{x}}) &= \frac{%
    \mathcal{N}(\bar{\bm{x}}|\boldsymbol{\mu}_{z},\widehat{\boldsymbol{\Sigma}})}{%
    \sum_{z'=1}^{|\mathcal{Z}|}\mathcal{N}(\bar{\bm{x}}|\boldsymbol{\mu}_{z'},\widehat{\boldsymbol{\Sigma}})}
  = \frac{%
    e^{-\frac{1}{2}\xi_{\widehat{\boldsymbol{\Sigma}}}(\bar{\bm{x}},\boldsymbol{\mu}_{z})}}{%
    \sum_{z'=1}^{|\mathcal{Z}|}e^{-\frac{1}{2}\xi_{\widehat{\boldsymbol{\Sigma}}}(\bar{\bm{x}},\boldsymbol{\mu}_{z'})}},
\end{align*}
from which the maximum a posteriori class estimate can be obtained as
\begin{equation*}
  \hat{z}=\argmax_{z\in\mathcal{Z}}p_{\mathrm{pred}}(z|\bar{\bm{x}}).
\end{equation*}

\subsubsection{General Model for CNN Classification}\label{sec:cnn_sensing_model}
For a more realistic but analytically intractable evaluation, we also consider a general CNN sensing model for nonlinear classification. Here, each feature vector $\bm{x}_i$ (including the query feature vector) is extracted from an input image $\tilde{x}_i$ using an encoder CNN as $\bm{x}_i=f_{\mathrm{enc}}(\tilde{x}_i)$\footnote{Although we assume that the query feature vector is constructed from an image using the same encoder CNN as the sensors, our proposed can also be applied to the case where the query feature vector is constructed using a different model and from a different data type, such as from a text query.}. We will assume that all sensors observing a given class observe different \emph{views} of the same object, e.g., from various angles or under different light conditions. For each class $z\in\mathcal{Z}$, this correlation is captured by the latent random variable $\kappa_z\sim p(\kappa_z|z)$ and implicitly described by the images in the dataset used for training and testing. More formally, we will assume that the input images $\tilde{\mathsf{X}}=(\tilde{x}_q,\tilde{x}_1,\ldots,\tilde{x}_M)$ are drawn from the distribution
\begin{equation}
  p(\tilde{\mathsf{X}}|\mathsf{Z})=p(\tilde{x}_q|z_q)\prod_{z=1}^{|\mathcal{Z}|}\int_{\kappa_z}\prod_{i\in\mathcal{M}_z}p(\tilde{x}_i|\kappa_z)p(\kappa_z|z)\,\mathrm{d}\kappa_z.\label{eq:cnn_feature_dist}
\end{equation}
Under this assumption, the set of feature vectors $\mathsf{X}=(f_{\mathrm{enc}}(\tilde{x}_q),f_{\mathrm{enc}}(\tilde{x}_1),\ldots,f_{\mathrm{enc}}(\tilde{x}_M))$ extracted from images $\tilde{\mathsf{X}}$ follows a distribution of the form in \cref{eq:feature_dist}.

For classification, the edge server employs a neural network that is composed of feature fusion followed by a classifier. The feature fusion learns the relevancy metric $\chi(\bm{x}_m,\bm{x}_q)$ and computes a fused feature vector $\bar{\bm{x}}$ as in \cref{eq:feature_fusion}. The fused feature vector $\bar{\bm{x}}$ is passed to a neural network classifier $f_{\mathrm{ser}}(\cdot)$, which outputs a $|\mathcal{Z}|$-dimensional vector through a softmax activation function, so that which each entry approximates $p_{\mathrm{pred}}(z|\bar{\bm{x}})$.
The class estimate $\hat{z}$ is then obtained as
\begin{align}
  \hat{z}=\argmax_{z\in\mathcal{Z}}\left[f_{\mathrm{ser}}(\bar{\bm{x}})\right]_{z},\label{eq:nn_pred}
\end{align}
where $\left[f_{\mathrm{ser}}(\bar{\bm{x}})\right]_{z}$ denotes the $z$-th entry of the vector $f_{\mathrm{ser}}(\bar{\bm{x}})$.

In this paper, we will not consider the specific architecture and quality of the feature encoder and classifier models $f_{\mathrm{enc}}(\cdot)$ and $f_{\mathrm{ser}}(\cdot)$, but instead assume that they are given, e.g., as pretrained models. The specific setup used in the numerical results is provided in \cref{sec:results}.

\subsection{Communication Model}

\begin{figure}
  \centering
  \includetikzimage<frame_model><font=\scriptsize>
  \caption{The considered frame structure containing semantic matching followed by random access observation transmission.}\label{fig:frame_model}
\end{figure}

\subsubsection{Semantic Matching (Downlink)}
In the semantic matching phase, the edge server constructs an $l$-dimensional semantic query vector $\bm{q}\in\mathbb{R}^l$ from its query feature vector $\bm{x}_q$. We assume that the dimension of the query vector is less than that of the query feature vector, i.e., $l\le d$. The query vector is transmitted to all devices with the purpose of allowing each device to compute a \emph{matching score} $\bar{\chi}(\bm{x}_m, \bm{q})\ge 0$ that approximates the relevancy of its sensed feature vector given the query feature vector, i.e.,
\begin{equation*}
  \bar{\chi}(\bm{x}_m, \bm{q})\approx \chi(\bm{x}_m, \bm{x}_q).
\end{equation*}
We assume that each entry of the query vector is quantized to sufficient number of bits, $\bar{Q}$, such that the quantization noise is negligible.
The resulting $l\bar{Q}$ bits are transmitted using $L_{\mathrm{dl}}$ (complex) symbols, resulting in a transmission rate of $R_{\mathrm{dl}}=l\bar{Q}/L_{\mathrm{dl}}$ bits/symbol.
We assume independent Rayleigh block-fading channels to each device, and for simplicity, we will assume that all devices have the same average signal-to-noise ratio (SNR) $\gamma$. This assumption allows us to focus on the subsequent uplink random access phase where performance is dominated by packet collisions rather than downlink channel variations.\footnote{In scenarios where users have different average SNRs, users with low SNRs would experience a higher query transmission failure probability, which would prevent them from performing matching, leading to non-homogeneous matching error probabilities. A rigorous treatment of this scenario would require a model that links the correlated observations to the SNRs, which we leave to future work.} The query transmission error probabilities experienced by the devices are then independent and given by the outage probability
\begin{equation*}
  p_{\mathrm{err}}^{\mathrm{dl}}=1-\exp\left(\frac{-\left(2^{R_{\mathrm{dl}}}-1\right)}{\gamma}\right).
\end{equation*}
The devices that successfully receive the query will perform threshold-based semantic matching, i.e., they will assume that their observation matches the query whenever $\bar{\chi}(\bm{x}_m,\bm{q})\ge \tau$ for some predefined threshold $\tau\ge 0$. We assume that devices that experience outage refrain from transmitting in the uplink.

\subsubsection{Random Access Observation Transmission (Uplink)}
The devices with matching observations transmit their observations to the server using IRSA over a shared slotted random access channel comprising $L_{\mathrm{ul}}$ transmission slots\footnote{We note that our scheme can be realized using any random access protocol, but we focus on IRSA due to its generality and practical relevance. Applying our scheme to other random access protocols is straightforward.}. In IRSA, each of the transmitting devices transmits a random number $L_m$ of identical packet replicas in slots selected uniformly at random without replacement among all $L_{\mathrm{ul}}$ slots. The number of replicas is drawn from the common degree distribution $\{\Lambda_{\ell}\}_{\ell=0}^{L_{\mathrm{ul}}}$, where $\Lambda_{\ell}=\Pr(L_m=\ell)$. Following the literature on IRSA (see, e.g.,~\cite{liva11irsa}), we will represent the degree distribution in compact polynomial form as $\Lambda(x)=\sum_{\ell=1}^{L_{\mathrm{ul}}} \Lambda_{\ell}x^{\ell}$. The degree distribution is assumed to be given.

At the end of the frame, the receiver decodes the packets iteratively using successive interference cancellation. Specifically, in each iteration, the receiver decodes packets in slots with only a single transmission, and then subtracts the remaining replicas of the decoded packets from their respective slots. In line with the IRSA literature, we assume that the locations of the replicas are revealed to the receiver after successful decoding of a packet, which can be achieved, e.g., by, including the seed used to generate the random replicas in the packet. The decoding and cancellation process is repeated until no slots with a single transmission remain, and any remaining transmissions are assumed to have failed. Note that $\Lambda(x)=x$ corresponds to conventional slotted ALOHA with a single replica. Averaging over the random replicas, the transmission failure probability is the same for all transmitting devices and denoted as $p_{\mathrm{err}}^{\mathrm{ul}}$.

\subsection{Performance Metrics}\label{sec:performance_metrics}
Our overall goal is to maximize the expected inference (or classification) accuracy defined as
\begin{align}
  \text{Inference accuracy} &= \frac{1}{|\mathcal{Z}|}\sum_{z_q=1}^{|\mathcal{Z}|}\E\left[\Pr(\hat{h}=z_q|z_q)\,\middle|\,z_q\right],\label{eq:accuracy}
\end{align}
where the expectation is over the observed feature vectors and the uplink transmission.

In addition to inference accuracy, we are also interested in the overall retrieval performance, characterized by the probability of MD and FA. The probability of MD, $\varepsilon_{\mathrm{MD}}$, is defined as the probability that a device that observes the query class is not among the set of received observations, while the probability of FA, $\varepsilon_{\mathrm{FA}}$, is defined as the probability that a device that does not observe the query class is among the set of received observations. Formally,
\begin{align}
  \varepsilon_{\mathrm{MD}}\!&=\!\frac{1}{|\mathcal{Z}|}\!\sum_{z_q=1}^{|\mathcal{Z}|}\E\!\left[\frac{1}{|\mathcal{M}_{z_q}|}\!\sum_{m_i\in \mathcal{M}_{z_q}}\!\Pr(m_i\!\notin\! \mathcal{M}_{\mathrm{rx}})\,\middle|\,z_q\right],\label{eq:p_md}\\
  \varepsilon_{\mathrm{FA}}\!&=\!\frac{1}{|\mathcal{Z}|}\!\sum_{z_q=1}^{|\mathcal{Z}|}\E\!\left[\frac{1}{|\mathcal{M}_{\mathrm{rx}}|}\!\sum_{m_i\in\mathcal{M}_{\mathrm{rx}}}\!\Pr(m_i\!\notin\! \mathcal{M}_{z_q})\,\middle|\,z_q\right],\label{eq:p_fa}
\end{align}
where both expectations are the sets $\mathcal{M}_{z_q}$ and $\mathcal{M}_{\mathrm{rx}}$.

\section{Semantic Matching Design}\label{sec:semantic_matching}
In this section, we present and analyze the proposed semantic matching phase for the GMM in details. We first describe the proposed query and matching design, which relies on linear projections, and then consider the problem of selecting the projection matrix.

\subsection{Query and Matching Design}
Recall that our aim is to construct a low-dimensional query vector $\bm{q}\in\mathbb{R}^l$ that allows devices to compute a matching score $\bar{\chi}(\bm{x}_m,\bm{q})$ that approximates the relevancy score in \cref{eq:gaussian_similarity_score}.
For analytical tractability, we will construct the query as a linear orthogonal projection of the query feature vector. First, the squared Mahalanobis distance in \cref{eq:squared_mahalan} can be equivalently written
\begin{align*}
  \xi_{\boldsymbol{\Sigma}}(\bm{x}_m,\bm{x}_q)=\|\boldsymbol{\Sigma}^{-\frac{1}{2}}\bm{x}_m-\boldsymbol{\Sigma}^{-\frac{1}{2}}\bm{x}_q\|_2^2.
\end{align*}
Let $\mathbf{P}_l\in\mathbb{R}^{l\times d}$ denote the matrix that defines the projection to the $l$-dimensional subspace satisfying $\mathbf{P}_l\mathbf{P}_l^T=\mathbf{I}_l$.
We then approximate the squared Mahalanobis distance as
\begin{align}
  \xi_{\boldsymbol{\Sigma}}(\bm{x}_m,\bm{x}_q)&\approx (d/l)\|\mathbf{P}_l\boldsymbol{\Sigma}^{-\frac{1}{2}}\bm{x}_m-\mathbf{P}_l\boldsymbol{\Sigma}^{-\frac{1}{2}}\bm{x}_q\|_2^2,\label{eq:approx_similarity_score}
\end{align}
where $(\sqrt{d/l})$ is a normalization factor to compensate for the reduced dimensionality. Note that the approximation in \cref{eq:approx_similarity_score} becomes an equality when $l=d$.

Under this approximation, we define the query vector $\bm{q}$ and a corresponding \emph{key vector} $\bm{k}_m\in\mathbb{R}^l$ as
\begin{align}
  \bm{q} &= (\sqrt{d/l})\mathbf{P}_l\boldsymbol{\Sigma}^{-\frac{1}{2}}\bm{x}_q,\label{eq:query}\\
  \bm{k}_m&=(\sqrt{d/l})\mathbf{P}_l\boldsymbol{\Sigma}^{-\frac{1}{2}}\bm{x}_m,\ m=1,\ldots,M.\label{eq:key}
\end{align}
Combining these definitions with \cref{eq:gaussian_similarity_score,eq:approx_similarity_score}, we obtain the matching score as
\begin{equation*}
  \bar{\chi}(\bm{x}_m, \bm{q}) = e^{-\frac{1}{d}\|\bm{k}_m-\bm{q}\|_2^2},
\end{equation*}
where for simplicity we omit the dependency of $\bm{x}_m$ on $\bm{k}_m$.
Note that $\bar{\chi}(\bm{x}_m, \bm{q})$ is a monotonically decreasing function of the approximate squared Mahalanobis distance, and thus also of the approximated relevancy of the device feature vector.

The following lemma characterizes the resulting probabilities of MD and FA of semantic matching prior to the uplink phase, denoted as $\varepsilon_{\mathrm{MD},\mathrm{match}}(\tau)$ and $\varepsilon_{\mathrm{FA},\mathrm{match}}(\tau)$, respectively.

\begin{lemma}\label{lemma:md_fa}
  For fixed $p_{\mathrm{err}}^{\mathrm{dl}}$ and for any threshold $0<\tau\le 1$, the probability of MD $\varepsilon_{\mathrm{MD},\mathrm{match}}(\tau)$ and FA $\varepsilon_{\mathrm{FA},\mathrm{match}}(\tau)$ in the semantic matching phase under the GMM are given as
  \begin{align}
    \varepsilon_{\mathrm{MD},\mathrm{match}}(\tau) &= 1-(1-p_{\mathrm{err}}^{\mathrm{dl}})\left(1-\frac{\Gamma\left(l/2,\tilde{\tau}/4\right)}{\Gamma(l/2)}\right),\label{eq:md}\\
    \varepsilon_{\mathrm{FA},\mathrm{match}}(\tau) &= \left(1-p_{\mathrm{err}}^{\mathrm{dl}}\right)\bigg[1-\frac{2}{|\mathcal{Z}|(|\mathcal{Z}|-1)}\nonumber\\
  &\quad\times\sum_{i=1}^{|\mathcal{Z}|}\sum_{j=i+1}^{|\mathcal{Z}|}Q_{\frac{l}{2}}\left(\sqrt{G_{i,j}(\mathbf{P}_l)/2}, \sqrt{\tilde{\tau}/2}\right)\bigg]\label{eq:fa},
\end{align}
where $\tilde{\tau}=l\ln(1/\tau)$ and
\begin{align*}
  G_{i,j}(\mathbf{P}_l)=\left(\boldsymbol{\mu}_i-\boldsymbol{\mu}_j\right)^T\boldsymbol{\Sigma}^{-\frac{1}{2}}\mathbf{P}_l^T\mathbf{P}_l\boldsymbol{\Sigma}^{-\frac{1}{2}}\left(\boldsymbol{\mu}_i-\boldsymbol{\mu}_j\right).
\end{align*}
\end{lemma}
\begin{IEEEproof}
  See \cref{app:proof_lemma_md_fa}.
\end{IEEEproof}

The MD probability in \cref{lemma:md_fa} is a monotonically increasing function in $\tau$, while the FA is decreasing in $\tau$, revealing the tradeoff between MD and FA controlled by $\tau$. Furthermore, the FA probability is a monotonically increasing function of $G_{i,j}(\mathbf{P}_l)$. This quantity is equivalent to the symmetric Kullback-Leibler (KL) divergence between the feature vector distributions of classes $i$ and $j$ in the reduced dimensionality space, and is often referred to as the \emph{discriminant gain} since it represents the discernibility between the two classes~\cite{lan23progressftx}. Note that only the probability of FA depends on $\mathbf{P}_l$. Thus, for a fixed dimension $l$, the projection matrix $\mathbf{P}_l$ should be chosen to minimize the probability of FA, which we do next.

\subsection{Query Projection Optimization}
In this section, we consider the problem of selecting the projection matrix $\mathbf{P}_l$ so as to minimize the probability of FA.
However, direct optimization of $\varepsilon_{\mathrm{FA},\mathrm{match}}(\tau)$ is challenging because of the non-linearity introduced by $Q_{M}(\alpha,\beta)$, which has no closed form. Furthermore, because of this non-linearity, the matrix $\mathbf{P}_l$ that minimizes $\varepsilon_{\mathrm{FA},\mathrm{match}}(\tau)$ depends on the choice of the matching threshold $\tau$. This means that the optimal $\mathbf{P}_l$ needs to be computed and communicated to the devices (or stored) for each particular choice of $\tau$, which may not be desirable in practice. Instead, we propose the more pragmatic strategy to pick the $\mathbf{P}_l$ that maximizes the average discriminant gain $G_{i,j}(\mathbf{P}_l)$ over all classes. This problem can be written:
\begin{equation}
\begin{aligned}
  \max_{\mathbf{P}_l\in\mathbb{R}^{l\times d}} \quad & \frac{2}{|\mathcal{Z}|(|\mathcal{Z}|-1)}\sum_{i=1}^{|\mathcal{Z}|}\sum_{j=i+1}^{|\mathcal{Z}|}G_{i,j}(\mathbf{P}_l),\\
  \textrm{s.t.} \quad & \mathbf{P}_l\mathbf{P}_l^T=\mathbf{I}_l.
\end{aligned}\label{eq:max_pl}
\end{equation}
Since $\varepsilon_{\mathrm{FA},\mathrm{match}}(\tau)$ is monotonically decreasing in the pairwise discriminant gains as discussed above, it can be expected that the solution to \cref{eq:max_pl} also performs well in terms of the FA probability across a wide range of thresholds $\tau$.
Furthermore, the solution to \cref{eq:max_pl} has a closed form, as stated in the following proposition.
\begin{proposition}\label{prop:query_selection}
  Define $\mathbf{W}\in\mathbb{R}^{d\times d}$ as
  \begin{equation*}
    \mathbf{W}=\sum_{i=1}^{|\mathcal{Z}|} \boldsymbol{\Sigma}^{-\frac{1}{2}}(\boldsymbol{\mu}_i-\bar{\boldsymbol{\mu}})(\boldsymbol{\mu}_i-\bar{\boldsymbol{\mu}})^T\boldsymbol{\Sigma}^{-\frac{1}{2}}
  \end{equation*}
  where $\bar{\boldsymbol{\mu}}=\frac{1}{|\mathcal{Z}|}\sum_{i=1}^{|\mathcal{Z}|}\boldsymbol{\mu}_i$, and let $\bm{v}_1,\bm{v}_2,\ldots,\bm{v}_d\in\mathbb{R}^d$ be the eigenvectors of $\mathbf{W}$ sorted in descending order by the magnitude of their corresponding eigenvalues.
  For any feature dimension $1\le l \le d$, the $\mathbf{P}_l$ that solves \cref{eq:max_pl} is
  \begin{equation*}
    \mathbf{P}_l = \begin{bmatrix}
      \bm{v}_1^T & \bm{v}_2^T & \cdots & \bm{v}_l^T
    \end{bmatrix}^T.
  \end{equation*}
\end{proposition}
\begin{IEEEproof}
  See \cref{app:proof_prop_query_selection}.
\end{IEEEproof}

\cref{prop:query_selection} is equivalent to applying Fisher's linear discriminant analysis (LDA)~\cite{bishop2006pattern}, which finds the linear projection that maximizes the ratio between the between-class and the within-class variances. This is intuitive, since the motivation behind LDA and that of maximizing the discriminant gain are essentially the same, namely to be able to distinguish samples drawn from distinct classes.

\section{Random Access Design}\label{sec:threshold}
In this section, we extend the previous analysis to include the random access phase. This phase introduces a new aspect to the tradeoff between MD and FA. Specifically, while a low matching threshold $\tau$ increases the probability of retrieving the desired observations, it may lead to a large number of collisions in the random access phase caused by FA. On the other hand, a high matching threshold reduces the risk of FA, but increases the probability of MD. To balance this tradeoff, we aim to optimize the matching score threshold $\tau$ to maximize the end-to-end classification accuracy by taking into account both semantic matching and collisions in the random access phase.

Unfortunately, direct optimization of classification accuracy is challenging as it depends on the distribution of the weighted feature vector, which is hard to compute. Instead, we consider the MD probability as a proxy for classification accuracy, and seek to minimize the MD probability. This is in turn equivalent to maximizing the number of received true positive (TP) observations, denoted as $N_{\mathrm{TP}}=|\mathcal{M}_{z_q}\cap\mathcal{M}_{\mathrm{rx}}|$. Using this, the considered threshold selection problem can be stated:
\begin{equation}
\begin{aligned}
  \max_{\tau\in[0,1]} \quad & \E\left[N_{\mathrm{TP}}\,\middle|\,\tau\right],\label{eq:tau_select}
\end{aligned}
\end{equation}
where the expectation is over the feature vectors, the downlink query transmission, and the transmission of the matching observations over the random access channel in the uplink.
Compared to maximizing the inference accuracy directly, maximizing the expected number of TP observations neglects the impact of FA on the inference accuracy. However, since FAs contribute to an increased collision probability during random access, maximizing the number of received TPs will indirectly suppress the number FAs. After receiving the observations, the weighted feature fusion (\cref{eq:feature_fusion}) will give weight to observations that provide a good match, which in turn suppresses the impact of FAs. Therefore, while the relationship between~\eqref{eq:tau_select} and \eqref{eq:accuracy}  is hard to characterize, the number of received TPs intuitively provides a good proxy for the inference accuracy.

Computing the objective function in \cref{eq:tau_select} is still intractable because of correlation among the matching observations caused by the common query feature vector. Therefore, we will seek an approximate solution by assuming that the devices observe independent query feature vectors drawn from the query class. In this case, the probability of a true positive transmission is the probability that a device observes the query class and its relevancy score is greater than $\tau$, which can be expressed using the result in \cref{lemma:md_fa} as
\begin{align*}
  p_{\mathrm{TP,tx}}(\tau) = p_{\mathrm{pos}}\left(1-\varepsilon_{\mathrm{MD},\mathrm{match}}(\tau)\right).
\end{align*}
Similarly, the probability of a FA transmission can be written
\begin{align*}
  p_{\mathrm{FA,tx}}(\tau) = (1-p_{\mathrm{pos}})\varepsilon_{\mathrm{FA},\mathrm{match}}(\tau).
\end{align*}
Under the assumption of independent queries introduced above, the total number of transmissions in the random access phase follows a binomial distribution with $M$ trials and success probability $p_{\mathrm{TP,tx}}(\tau)+p_{\mathrm{FA,tx}}(\tau)$. For large $M$, which is the regime of interest, this number can be accurately approximated as a Poisson distribution with rate $\bar{\lambda}(\tau)=M(p_{\mathrm{TP,tx}}(\tau)+p_{\mathrm{FA,tx}}(\tau))$. Furthermore, under this assumption we may write the number of TP transmissions and the number of FA transmissions as independent Poisson distributed random variables with rates
\begin{align}
  \bar{\lambda}_{\mathrm{TP}}(\tau)&=Mp_{\mathrm{TP,tx}}(\tau),\label{eq:bar_lambda_td}\\
  \bar{\lambda}_{\mathrm{FA}}(\tau)&=Mp_{\mathrm{FA,tx}}(\tau)\label{eq:bar_lambda_fa},
\end{align}
respectively, such that $\bar{\lambda}(\tau)=\bar{\lambda}_{\mathrm{TP}}(\tau)+\bar{\lambda}_{\mathrm{FA}}(\tau)$.

In the following, we first present the analysis of conventional slotted ALOHA (i.e., $\Lambda(x)=x$), for which a simple approximate solution can be obtained. We will then generalize the result to the case of IRSA, which requires more involved approximations.

\subsection{Conventional Slotted ALOHA}\label{sec:slotted_aloha_analysis}
In conventional slotted ALOHA, transmission errors occur if another device transmits in the same slot. For Poisson arrivals and $L_{\mathrm{ul}}$ slots, this happens with probability
\begin{align}
  p_{\mathrm{err}}^{\mathrm{ul}}(\tau)=1-e^{-\bar{\lambda}(\tau)/L_{\mathrm{ul}}}.\label{eq:p_err_ul_aloha}
\end{align}
The number of received true positive observations can then be approximated as
\begin{align}
  \E\left[N_{\mathrm{TP}}\,\middle|\,\tau\right]\approx \bar{\lambda}_{\mathrm{TP}}(\tau)\left(1-p_{\mathrm{err}}^{\mathrm{ul}}(\tau)\right),\label{eq:approx_ntp}
\end{align}
where the approximation comes from the assumption that devices receive independent queries and from the Poisson approximation.
The optimal $\tau$ that maximizes the right-hand side of \cref{eq:approx_ntp} is presented in the following proposition.
\begin{proposition}\label{prop:opt_tau}
  For $l\ge 1$ and $L_{\mathrm{ul}}>0$, there exists a unique threshold $0<\tau\le 1$ that maximizes the expected number of received true positive observations under the simplifying assumptions defined above. Furthermore, defining $\tilde{\tau}=l\ln(1/\tau)$, the optimal $\tau$ is the solution to
  \begin{multline}
    L_{\mathrm{ul}}-M(1-p_{\mathrm{err}}^{\mathrm{dl}})\gamma\left(\frac{l}{2},\frac{\tilde{\tau}}{4}\right)\\
  \times\left(p_{\mathrm{pos}}+\frac{(1-p_{\mathrm{pos}})2^{l-1}\Gamma(l/2)}{|\mathcal{Z}|(|\mathcal{Z}|-1)}\phi\left(\tilde{\tau}\right)\right)=0,\label{eq:opt_tau}
  \end{multline}
  where
  \begin{multline*}
    \phi\left(\tilde{\tau}\right)=\tilde{\tau}^{1/2-l/4}
    \sum_{i=1}^{|\mathcal{Z}|}\sum_{j=i+1}^{|\mathcal{Z}|}\bigg[\frac{e^{-\frac{1}{4}G_{i,j}(\mathbf{P}_l)}}{G_{i,j}(\mathbf{P}_l)^{l/4-1/2}}\\
    \times I_{l/2-1}\left(\frac{1}{2}\sqrt{G_{i,j}(\mathbf{P}_l)\tilde{\tau}}\right)\bigg],
  \end{multline*}
  and $\gamma(s,x)$ and $I_{n}(x)$ are the lower incomplete gamma function and the modified Bessel function of the first kind, respectively.
\end{proposition}
\begin{IEEEproof}
  See \cref{app:proof_prop_opt_tau}.
\end{IEEEproof}

The left-hand side of \cref{eq:opt_tau} is an increasing function of $\tau$, and thus the optimal $\tau$ can be solved efficiently using standard root-finding algorithms, such as bisection search. Once the optimal $\tau$ has been obtained, the probability of MD and FA as defined in \cref{eq:p_md,eq:p_fa} can be approximated by combining the results in \cref{lemma:md_fa} with \cref{eq:p_err_ul_aloha}:
\begin{align}
  \varepsilon_{\mathrm{MD}}&\approx \varepsilon_{\mathrm{MD},\mathrm{match}}(\tau)+p_{\mathrm{err}}^{\mathrm{ul}}(\tau)\left(1-\varepsilon_{\mathrm{MD},\mathrm{match}}(\tau)\right),\label{eq:slottedaloha_md}\\
  \varepsilon_{\mathrm{FA}}&\approx \left(1-p_{\mathrm{err}}^{\mathrm{ul}}(\tau)\right)\varepsilon_{\mathrm{FA},\mathrm{match}}(\tau).\label{eq:slottedaloha_fa}
\end{align}
Note that, in addition to the GMM parameters, the optimal threshold depends on $p_{\mathrm{pos}}$ and $p_{\mathrm{err}}^{\mathrm{ul}}$, which may change over time and depending on the specific task. In such case, the optimal threshold needs to be recomputed and can be broadcast along with the query (at the cost of a few bytes).

\subsection{Irregular Repetition Slotted ALOHA}\label{sec:opt_tau_irsa}
While the performance of conventional ALOHA is severely limited by collisions, IRSA can support a much larger number of devices with the same number of slots by allowing multiple repetitions and successive interference cancellation~\cite{liva11irsa}. However, the analysis of IRSA is more involved, and the error probability cannot be expressed in closed-form. Instead, we will resort to approximations proposed in~\cite{ivanov17slottedaloha,amat18aloha}, which have been applied with success to a wide range of scenarios, e.g.,~\cite{munari21aloha,ngo21aloha}. These approximations rely on the fact that the error probability exhibits an error-floor regime for small frame loads (defined as $\bar{\lambda}(\tau)/L_{\mathrm{ul}}$), and a waterfall regime for large frame loads, which can be approximated separately. As in~\cite{munari21aloha}, we will approximate the total error probability as the sum of the approximations in the two regimes, i.e.,
\begin{equation}
  p_{\mathrm{err}}^{\mathrm{ul}}(\tau)\approx p_{\mathrm{err},\mathrm{ef}}^{\mathrm{ul}}(\tau)+p_{\mathrm{err},\mathrm{wf}}^{\mathrm{ul}}(\tau),\label{eq:p_err_irsa}
\end{equation}
where $p_{\mathrm{err},\mathrm{ef}}^{\mathrm{ul}}(\tau)$ and $p_{\mathrm{err},\mathrm{wf}}^{\mathrm{ul}}(\tau)$ are approximations of the error probability in the error-floor and in the waterfall regions, respectively. The waterfall approximation is~\cite{amat18aloha}
\begin{align*}
  p_{\mathrm{err},\mathrm{wf}}^{\mathrm{ul}}(\tau)=\alpha_1 Q\left( \frac{\sqrt{L_{\mathrm{ul}}}\left(\alpha_2-\alpha_3 L_{\mathrm{ul}}^{-2/3}-\bar{\lambda}(\tau)/L_{\mathrm{ul}}\right)}{\sqrt{\alpha_0^2+\bar{\lambda}(\tau)/L_{\mathrm{ul}}}} \right),
\end{align*}
where $\alpha_0,\alpha_1,\alpha_2,\alpha_3\in\mathbb{R}$ are constants that depend on the specific degree distribution $\Lambda(x)$ and obtained through numerical analysis of the decoding process. The error probability in the error-floor regime is approximated as~\cite{ivanov17slottedaloha} 
\begin{align}
  &p_{\mathrm{err},\mathrm{ef}}^{\mathrm{ul}}(\tau)\nonumber\\
  &\!=\! \sum_{s=1}^A \phi(s,\tau)\boldsymbol{\nu}[s]\boldsymbol{\beta}_0[s]\binom{L_{\mathrm{ul}}}{\boldsymbol{\beta}_1[s]}\!\prod_{\ell=1}^{L_{\mathrm{ul}}}\frac{\Lambda_{\ell}^{\boldsymbol{\nu}_{\ell}[s]}}{\boldsymbol{\nu}_{\ell}[s]!}\binom{L_{\mathrm{ul}}}{\ell}^{-\boldsymbol{\nu}_{\ell}[s]},\label{eq:p_err_ef}
\end{align}
where
\begin{align*}
  \phi(s,\tau) = \sum_{k=0}^{\boldsymbol{\nu}[s]-1} (-1)^{\boldsymbol{\nu}[s]-1+k}\bar{\lambda}(\tau)^k\frac{(\boldsymbol{\nu}[s]-1)!}{k!}.
\end{align*}
The constants $A\in\mathbb{Z}$, $\boldsymbol{\nu}_{\ell}\in\mathbb{R}^{|\mathcal{A}|},\ell=1,2,\ldots,L_{\mathrm{ul}}$, as well as $\boldsymbol{\beta}_0,\boldsymbol{\beta}_1\in\mathbb{R}^{|\mathcal{A}|}$ depend on the specific degree distribution, and  $\boldsymbol{\nu}=\sum_{\ell=1}^{L_{\mathrm{ul}}}\boldsymbol{\nu}_{\ell}$.
For instance, for $\Lambda(x)=x^3$ \cref{eq:p_err_ef} simplifies to
\begin{align*}
  p_{\mathrm{err},\mathrm{ef}}^{\mathrm{ul}}(\tau) = \sum_{s=1}^2 \frac{\phi(s,\tau)\boldsymbol{\nu}[s]\boldsymbol{\beta}_0[s]}{\boldsymbol{\nu}[s]!}\binom{L_{\mathrm{ul}}}{\boldsymbol{\beta}_1[s]}\binom{L_{\mathrm{ul}}}{3}^{-\boldsymbol{\nu}[s]},
\end{align*}
and the parameters are given as $\alpha_0=0.497867$, $\alpha_1=0.784399$, $\alpha_2=0.818469$, $\alpha_3=0.964528$, $A=2$, $\boldsymbol{\nu}=(2,3)$, $\boldsymbol{\beta}_0=(1,24)$, $\boldsymbol{\beta}_1=(3,4)$~\cite{munari21aloha}. The resulting approximation is compared to simulations for $L_{\mathrm{ul}}\in\{25,50,100\}$ in \cref{fig:irsa_approximation} under the assumption of Poisson arrivals, showing that the approximation closely captures the trend of the error probability.

\begin{figure}
  \centering
  \includetikzimage<irsa_approximation><font=\footnotesize>
  \caption{Comparison between simulated error probability and approximation \eqref{eq:p_err_irsa} for $\Lambda(x)=x^3$ under Poisson arrivals.}\label{fig:irsa_approximation}
\end{figure}

We can now approximate the expected number of received true positive observations by using \cref{eq:p_err_irsa} in \cref{eq:approx_ntp}. The resulting approximation is accurate in the interval $\bar{\lambda}(\tau)\in[0,L_{\mathrm{ul}}]$. For $\bar{\lambda}(\tau)>L_{\mathrm{ul}}$, the error probability is close to $1$, and thus $\E\left[N_{\mathrm{TP}}\,\middle|\,\tau\right]$ is approximately zero. The threshold optimization problem can then be written
\begin{equation}
\begin{aligned}
\max_{\tau\in [0,1]} \quad & \bar{\lambda}_{\mathrm{TP}}(\tau)\left(1-p_{\mathrm{err}}^{\mathrm{ul}}(\tau)\right)\\
\textrm{s.t.} \quad & \bar{\lambda}(\tau) < L_{\mathrm{ul}}.\label{eq:approx_ntp_irsa_opt}
\end{aligned}
\end{equation}

Solving \cref{eq:approx_ntp_irsa_opt} analytically is hard because of the complexity of the approximation of $p_{\mathrm{err}}^{\mathrm{ul}}(\tau)$, and we resort to numerical optimization. To this end, we note that since $\bar{\lambda}(\tau)$ is monotonically decreasing in $\tau$, the constraint can be equivalently written as $\tau_{\mathrm{lb}}\le \tau\le 1$, where $\tau_{\mathrm{lb}}=\inf_{\tau\in[0,1]}\{\tau | \bar{\lambda}(\tau)\le L_{\mathrm{ul}}\}$. Once $\tau_{\mathrm{lb}}$ has been determined, we maximize the objective function over $\tau\in[\tau_{\mathrm{lb}},1]$. However, since the objective function in \cref{eq:approx_ntp_irsa_opt} is not guaranteed to be convex, the optimization is sensitive to the initialization point. We address this issue by performing the optimization procedure multiple times with initial values of $\tau$ that are spaced at regular intervals over its domain, and then pick the best result across all initializations. While this generally does not guarantee to provide the globally optimal solution, experimental results suggest that the global optimum is always found even with only 10 initial values.

As with conventional ALOHA, the resulting MD and FA probabilities can be approximated using \cref{eq:slottedaloha_md,eq:slottedaloha_fa} with the approximation in \cref{eq:p_err_irsa}.

\section{CNN-Based Semantic Data Sourcing}\label{sec:neuralnet}
The previous sections studied the semantic query and matching problem under the GMM. In this section, we demonstrate how the idea can be applied to the more realistic and general CNN-based sensing model. As in the GMM case, we propose to solve the matching problem by comparing the query broadcast by the edge server to a local key constructed by each sensor. However, instead of designing the query and keys by hand for a specific relevancy function, we compute them using CNNs that we train to facilitate the end objective of maximizing the inference accuracy defined in \cref{eq:accuracy}. Note that the relevancy function is implicitly learned as part of the training. In the following, we first outline the query and key construction, and then explain how to optimize the matching score threshold.

\subsection{Attention-Based Semantic Matching}
As mentioned, the feature fusion step defined in \cref{eq:feature_fusion} resembles an attention mechanism, where the relevancy function gives the attention weights. Inspired by this, we propose to define the relevancy function inspired by the dot-product attention mechanism widely used in the literature~\cite{vaswani17attention}.
Recall that the query and key feature vectors $\bm{x}_q=f_{\mathrm{enc}}(\tilde{x}_q)$ and $\bm{x}_m=f_{\mathrm{enc}}(\tilde{x}_m),\,m=1,\ldots,M$ are generated by passing the observed objects through a CNN feature extractor $f_{\mathrm{enc}}(\cdot)$. We define the relevancy of a key feature vector given a query as
\begin{equation*}
  \chi(\bm{x}_m,\bm{x}_q) = e^{\frac{1}{l}\bm{q}^T\bm{k}_m}
\end{equation*}
where the query and key vectors $\bm{q},\bm{k}_m\in\mathbb{R}^l$ are generated from their corresponding feature vectors $\bm{x}_q$ and $\bm{x}_m$ by two feed-forward neural networks:
\begin{align*}
  \bm{q} &= f_{\mathrm{q}}(\bm{x}_q),\\
  \bm{k}_m&=f_{\mathrm{k}}(\bm{x}_m),\ m=1,\ldots,M.
\end{align*}
With this definition, the feature fusion step in \cref{eq:feature_fusion} is equivalent to a dot-product attention mechanism applied to the feature vectors.

The definitions above enable us to jointly learn the feature extractor $f_{\mathrm{enc}}(\cdot)$, the query and key encoders $f_{\mathrm{q}}(\cdot)$ and $f_{\mathrm{k}}(\cdot)$, and the classifier $f_{\mathrm{ser}}(\cdot)$ in an end-to-end fashion. To this end, we assume the availability of a multi-view dataset $\mathcal{D}_{\mathrm{train}}=\{(\tilde{\mathsf{X}}^{(n)},\mathsf{Z}^{(n)})\}_{n=1}^{N_\mathrm{train}}$ containing $N_{\mathrm{train}}$ examples, each comprising $M+1$ observations $\tilde{\mathsf{X}}^{(n)}=(\tilde{x}_q^{(n)},\tilde{x}_1^{(n)},\ldots,\tilde{x}_M^{(n)})$ of classes $\mathsf{Z}^{(n)}=(z_q^{(n)},z_1^{(n)},\ldots,z_M^{(n)})$ drawn according to the distribution in \cref{eq:cnn_feature_dist}. We employ a centralized learning and decentralized execution strategy commonly employed in distributed multi-agent settings~\cite{zhang21multiagent}. Specifically, during training we ignore the random access channel and assume that the edge server has access to the query example $\tilde{x}_q$ and all device observations $\tilde{x}_1,\ldots,\tilde{x}_M$, so that it can compute the query feature vector $\bm{x}_q$ and the device feature vectors $\bm{x}_m$ for all devices $m=1,\ldots,M$. Using these, we compute the fused feature vector $\bar{\bm{x}}$, which we use as input to the classifier to obtain the prediction as defined in \cref{eq:nn_pred}. The networks are then trained to minimize the end-to-end cross-entropy loss.
In the inference phase, we obtain the matching score function by normalizing the learned relevancy score as
\begin{equation*}
  \bar{\chi}(\bm{x}_m,\bm{q})=\frac{\chi(\bm{x}_m,\bm{x}_q)}{1+\chi(\bm{x}_m,\bm{x}_q)} = \frac{e^{\frac{1}{l}\bm{q}^T\bm{k}_m}}{1+e^{\frac{1}{l}\bm{q}^T\bm{k}_m}},
\end{equation*}
i.e., the sigmoid function applied to $(1/l)\bm{q}^T\bm{k}_m$. The feature fusion is based only on the feature vectors received after the random access phase as in \cref{eq:feature_fusion}.

Intuitively, it is expected that the neural network learns to weight feature vectors that contribute significantly towards a correct classification higher than less important feature vectors in the feature fusion step. Since we use this weight directly as the matching score, we expect that the matching score will reflect the importance of an observation given the query. However, it should be noted that the attention weights used for feature fusion in \cref{eq:feature_fusion} depend on the \emph{relative} relevancy scores, whereas the devices only have access to their own matching score. As a result, a large matching score does not, in general, imply that the feature will have a large weight in the fusion step.

\subsection{Matching Threshold Optimization}\label{sec:cnn_threshold_opt}
We estimate the probability of MD and FA matches from the empirical distribution functions. Given a calibration dataset $\mathcal{D}_{\mathrm{cal}}=\{(\tilde{\mathsf{X}}^{(n)},\mathsf{Z}^{(n)})\}_{n=1}^{N_{\mathrm{cal}}}$ structured as the training dataset $\mathcal{D}_{\mathrm{train}}$, the conditional empirical distributions can be computed by counting the number of MD and FA events across the dataset
\begin{align*}
  &\varepsilon_{\mathrm{MD},\mathrm{match}}'(\tau)\nonumber\\
  &\quad=\frac{1}{N_{\mathrm{cal}}}\sum_{n=1}^{N_{\mathrm{cal}}}\sum_{m=1}^M\frac{\mathbbm{1}\left[\bar{\chi}(\bm{x}_m^{(n)},\bm{q}^{(n)})< \tau\right]\mathbbm{1}\left[z_q^{(n)}=z_m^{(n)}\right]}{\sum_{m=1}^M \mathbbm{1}\left[z_q^{(n)}= z_m^{(n)}\right]},\\
  &\varepsilon_{\mathrm{FA},\mathrm{match}}'(\tau)\nonumber\\
&\quad=\frac{1}{N_{\mathrm{cal}}}\sum_{n=1}^{N_{\mathrm{cal}}}\sum_{m=1}^M\frac{\mathbbm{1}\left[\bar{\chi}(\bm{x}_m^{(n)},\bm{q}^{(n)})\ge \tau\right]\mathbbm{1}\left[z_q^{(n)}\neq z_m^{(n)}\right]}{\sum_{m=1}^M \mathbbm{1}\left[z_q^{(n)}\neq z_m^{(n)}\right]},
\end{align*}
where the query $\bm{q}^{(n)}$ and the device feature vector $\bm{x}_m^{(n)}$ are extracted using the CNN from $\tilde{x}_q^{(n)}$ and $\tilde{x}_m^{(n)}$, respectively.
By including the effect of the downlink transmission error probability $p_{\mathrm{err}}^{\mathrm{dl}}$, we obtain the marginal empirical distributions corresponding to the ones presented in \cref{lemma:md_fa} for the GMM:
\begin{align*}
  \varepsilon_{\mathrm{MD},\mathrm{match}}(\tau)&=p_{\mathrm{err}}^{\mathrm{dl}}+(1-p_{\mathrm{err}}^{\mathrm{dl}})\varepsilon_{\mathrm{MD},\mathrm{match}}'(\tau),\\
  \varepsilon_{\mathrm{FA},\mathrm{match}}(\tau)&=(1-p_{\mathrm{err}}^{\mathrm{dl}})\varepsilon_{\mathrm{FA},\mathrm{match}}'(\tau).
\end{align*}
These distributions allow us to approximate $\bar{\lambda}_{\mathrm{TP}}(\tau)$ and $\bar{\lambda}_{\mathrm{FA}}(\tau)$ using \cref{eq:bar_lambda_td,eq:bar_lambda_fa}, which in turn can be used to find the matching score threshold $\tau$ that maximizes the expected number of received true positive observations. Specifically, for conventional slotted ALOHA we optimize \cref{eq:approx_ntp}, while the threshold for IRSA is obtained by optimizing \cref{eq:approx_ntp_irsa_opt}. In both cases, we optimize the quantities numerically using the approach outlined for IRSA in \cref{sec:opt_tau_irsa}.

\section{Numerical Results}\label{sec:results}
In this section, we evaluate the proposed semantic data sourcing techniques and present their performance. We first consider the GMM sensing model, and then evaluate the machine learning model. In both scenarios, we compare the results to two benchmark schemes. The first benchmark, termed \emph{query-free access}, is traditional random access, in which each device transmits independently of the query with a predefined probability chosen to maximize the probability of successful transmission. The second benchmark, referred to as \emph{perfect matching}, is our proposed semantic query scheme but with perfect matching, i.e., without any FA but including downlink and random access communication errors. In the perfect matching benchmark, the threshold is optimized using exhaustive search to yield the best performance on the specific target statistic (e.g., to minimize the probability of MD or to maximize inference accuracy).

\subsection{Linear Classification using GMM}
We consider an observation model with $|\mathcal{Z}|=21$ classes and feature dimension $d=75$, in which the $j$-th entry of the $i$-th class centroid is given as $\boldsymbol{\mu}_i(j)=-1$ for $\lfloor d(i-1)/|\mathcal{Z}|\rfloor<j\le \lfloor di/|\mathcal{Z}|\rfloor$ and otherwise $\boldsymbol{\mu}_i(j)=1$. The entries of the covariance matrix $\boldsymbol{\Sigma}$ are assumed to be identical, i.e., $C_{j,j}=C$ for all $1\le j\le d$, where $C$ is chosen to achieve a desired average discriminant gain, $\overline{G}$, defined as
\begin{align*}
  \overline{G}=\frac{2}{|\mathcal{Z}|(|\mathcal{Z}|-1)}\sum_{i=1}^{|\mathcal{Z}|}\sum_{j=i+1}^{|\mathcal{Z}|}%
  \left(\boldsymbol{\mu}_i-\boldsymbol{\mu}_j\right)^T\boldsymbol{\Sigma}^{-1}\left(\boldsymbol{\mu}_i-\boldsymbol{\mu}_j\right).
\end{align*}
We assume a total of $M=200$ devices and the observation model described in \cref{sec:gaussian_sensing_model}, where each device observes the query class independently with probability $p_{\mathrm{pos}}=0.1$. The number of downlink symbols $L_{\mathrm{dl}}$ is chosen such that the error probability is $p_{\mathrm{err}}^{\mathrm{dl}}=0.1$, while the number of random access transmission slots in the uplink is $L_{\mathrm{ul}}=10$. Unless specified, we consider a query dimension of $l=20$. Quantizing each entry to $\overline{Q}=16$ bits for negligible distortion thus results in a query size of $40$ bytes. For IRSA, we assume the degree distribution $\Lambda(x)=x^3$.

\begin{figure}
  \centering
  \subfloat[MD, ALOHA]{\includetikzimage<results_gaussian_tau_vs_md><font=\footnotesize,trim axis right>\label{fig:gaussian_res_md_aloha}}\hfill
  \subfloat[MD, IRSA, $\Lambda(x)\!=\!x^3$]{\includetikzimage<results_gaussian_tau_vs_md_irsa><font=\footnotesize,trim axis left>\label{fig:gaussian_res_md_irsa}}\\\vspace{-1em}
  \subfloat[FA, ALOHA]{\includetikzimage<results_gaussian_tau_vs_fa><font=\footnotesize,trim axis right>\label{fig:gaussian_res_fa_aloha}}\hfill
  \subfloat[FA, IRSA, $\Lambda(x)\!=\!x^3$]{\includetikzimage<results_gaussian_tau_vs_fa_irsa><font=\footnotesize,trim axis left>\label{fig:gaussian_res_fa_irsa}}
  \caption{Probability of MD (\protect\subref{fig:gaussian_res_md_aloha}-\protect\subref{fig:gaussian_res_md_irsa}) and FA (\protect\subref{fig:gaussian_res_fa_aloha}-\protect\subref{fig:gaussian_res_fa_irsa}) vs. matching threshold $\tau$ for ALOHA and IRSA under the Gaussian observation model. Circles denote performance at the optimized $\tau$ found using the procedure in \cref{sec:threshold} by maximizing received TPs as a proxy for inference accuracy.}\label{fig:gaussian_res_md_fa}
\end{figure}

The probability of MD is shown in \cref{fig:gaussian_res_md_aloha,fig:gaussian_res_md_irsa} for $\overline{G}=10,20,40,60$ along with the approximation in \cref{eq:slottedaloha_md} obtained using the quantities derived in \cref{sec:slotted_aloha_analysis,sec:opt_tau_irsa}. Generally, a larger discriminant gain $\overline{G}$ leads to a lower probability of MD, approaching the minimum attainable MD under perfect matching (i.e., no FA) indicated by the blue dashed line. On the other hand, when $\overline{G}$ is small false positives significantly impact the probability of MD, approaching the performance of classical random access with independent activation probabilities as indicated by the green dotted line. Note that, while the semantic query scheme with a sub-optimal threshold has a higher MD probability than optimized classical random access, this comparison ignores the smaller performance gap that would be obtained if both were sub-optimally configured, as would be the case under, e.g., wrong model assumptions.
\cref{fig:gaussian_res_md_fa} also shows that the probability of MD with IRSA is lower than that of ALOHA, highlighting the advantage of a more efficient random access protocol. Furthermore, the analytical approximations accurately describe the performance in ALOHA, but are less tight in IRSA, especially for low values of $\tau$. As discussed in \cref{sec:opt_tau_irsa}, this is because the approximation is inaccurate when the frame load exceeds one, i.e., $\bar{\lambda}(\tau)/L_{\mathrm{ul}}\ge 1$. Nevertheless, the optimized thresholds, indicated by the circles, suggest that its accuracy is sufficient to obtain thresholds that approximately maximize the expected number of true positives, which is equivalent to minimizing the probability of MD.

The probability of FA is shown in \cref{fig:gaussian_res_fa_aloha,fig:gaussian_res_fa_irsa}. As for MD, a large discriminant gain generally results in a lower FA probability. While the difference in the FA probability between ALOHA and IRSA is smaller than for MD, the curves for IRSA peak lower than those of ALOHA. This is because ALOHA performs better than IRSA for high frame loads (i.e., low thresholds), where the fraction of FA transmissions is large. On the other hand, IRSA supports a much higher throughput than ALOHA at moderate frame loads, where the fraction of FA is lower. Finally, the figure shows that the analytical approximation deviates significantly from the simulated results.
The deviation stems primarily from modeling the device activations as an independent Poisson process, whereas the actual traffic is correlated, since all devices receive the same query, and the total number of devices ($M$) is finite. This discrepancy is most pronounced in the high-load regime (low $\tau$), where FAs are more likely to occur. Furthermore, the deviation is larger for IRSA (\cref{fig:gaussian_res_fa_irsa}) than for ALOHA (\cref{fig:gaussian_res_fa_aloha}) because of the approximation used to calculate the uplink error probability in IRSA. On the other hand, the uplink error probability in the case of ALOHA is accurate under the Poisson assumption. Nevertheless, the absolute approximation error is similar to that of the MD in \cref{fig:gaussian_res_md_aloha,fig:gaussian_res_md_irsa}, and the model still characterizes the main trend of the FA probabilities for both ALOHA and IRSA, providing a good approximation for optimization of $\tau$. As a result, the proposed semantic data sourcing method significantly reduces false alarms compared to what can be achieved with traditional random access, as indicated by the green dotted line.

\begin{figure}
  \centering
  \subfloat[ALOHA]{\includetikzimage<results_gaussian_tau_vs_acc><font=\footnotesize,trim axis right>\label{fig:gaussian_res_inf_acc_aloha}}\hfill
  \subfloat[IRSA, $\Lambda(x)=x^3$]{\includetikzimage<results_gaussian_tau_vs_acc_irsa><font=\footnotesize,trim axis left>\label{fig:gaussian_res_inf_acc_irsa}}
  \caption{Inference accuracy vs. matching threshold $\tau$ for \protect\subref{fig:gaussian_res_inf_acc_aloha} ALOHA and \protect\subref{fig:gaussian_res_inf_acc_irsa} IRSA under the Gaussian observation model. Circles denote performance at the optimized $\tau$ found using the procedure in \cref{sec:threshold} by maximizing received TPs as a proxy for inference accuracy.}\label{fig:gaussian_res_inf_acc}
\end{figure}

\cref{fig:gaussian_res_inf_acc} shows the end-to-end inference accuracy vs.\ the matching threshold. As expected, a larger discriminant gain results in higher accuracy. Furthermore, the optimized threshold approximately maximizes the accuracy, confirming that the number of received true positives serves as a good proxy for inference accuracy. However, despite the fact that the protocol with IRSA has a much smaller probability of MD (see \cref{fig:gaussian_res_md_aloha,fig:gaussian_res_md_irsa}), the inference accuracy achieved by IRSA and ALOHA is similar. This suggests that only a small number of true positives are sufficient to perform accurate inference in the specific task. Furthermore, for low discriminant gains the inference errors are dominated by ``noisy'' query examples and resulting FA observations, which cannot be compensated by the higher transmission success probability offered by IRSA. The proposed scheme significantly outperforms traditional query-free access (shown only for $\overline{G}=60$), highlighting the advantage of query-based access.

\begin{figure}
  \centering
  \subfloat[MD, ALOHA]{\includetikzimage<results_gaussian_p_pos_vs_md_aloha><font=\footnotesize,trim axis right>\label{fig:gaussian_res_p_pos_md_aloha}}\hfill
  \subfloat[MD, IRSA]{\includetikzimage<results_gaussian_p_pos_vs_md_irsa><font=\footnotesize,trim axis left>\label{fig:gaussian_res_p_pos_md_irsa}}\\\vspace{-1em}
  \subfloat[Inf. accuracy, ALOHA]{\includetikzimage<results_gaussian_p_pos_vs_acc_aloha><font=\footnotesize,trim axis right>\label{fig:gaussian_res_p_pos_acc_aloha}}\hfill
    \subfloat[Inf. accuracy, IRSA]{\includetikzimage<results_gaussian_p_pos_vs_acc_irsa><font=\footnotesize,trim axis left>\label{fig:gaussian_res_p_pos_acc_irsa}}
  \caption{Probability of MD (\protect\subref{fig:gaussian_res_p_pos_md_aloha}-\protect\subref{fig:gaussian_res_p_pos_md_irsa}) and inference accuracy (\protect\subref{fig:gaussian_res_p_pos_acc_aloha}-\protect\subref{fig:gaussian_res_p_pos_acc_irsa}) vs. probability of observing the query, $p_{\text{pos}}$, under the Gaussian observation model and using optimized thresholds.}\label{fig:gaussian_res_p_pos}
\end{figure}

To study how the performance is affected by the fraction of devices observing the query class, \cref{fig:gaussian_res_p_pos} shows the probability of MD and the inference accuracy vs.\ the probability that a device observes the query class, $p_{\mathrm{pos}}$, for optimized matching thresholds $\tau$. Generally, the probability of MD (\cref{fig:gaussian_res_p_pos_md_aloha,fig:gaussian_res_p_pos_md_irsa}) increases with $p_{\text{pos}}$, as the number of devices observing the query class gets larger than what can be supported by the random access channel. The curve is flat for small values of $p_{\text{pos}}$, which is caused by channel errors in both downlink and uplink. Specifically, the uplink error is caused by the fact that, as $p_{\text{pos}}$ decreases, it gets increasingly difficult to retrieve the true positive observations without also matching many FAs, which impacts the uplink error probability. This problem is absent from the perfect matching benchmark, for which the floor is caused only by the downlink error probability, i.e., $\varepsilon_{\text{MD}}\to p_{\text{err}}^{\text{dl}}=0.2$ as $p_{\text{pos}}\to 0$.
  Although the probability of missed detection increases with $p_{\text{pos}}$, the inference accuracy generally increases since the number of received true positive observations increases (\cref{fig:gaussian_res_p_pos_acc_aloha,fig:gaussian_res_p_pos_acc_irsa}). However, for large $p_{\text{pos}}$, the inference accuracy of the proposed method decreases, especially in the IRSA scenario. This is because the high threshold required to avoid an excessive number of transmissions comes at the cost of making the matching very sensitive to noisy queries. Specifically, there will be no or very few matches when the query is noisy, which occasionally results in low inference accuracies. This effect is more pronounced for the IRSA protocol than for ALOHA. This is because, while IRSA supports a higher throughput than ALOHA, it also has a higher error probability in the congested regime due to the packet repetitions. This lower error probability of ALOHA makes it possible to use a lower threshold, which in turn makes it less sensitive to query noise.
  However, this regime where $p_{\text{pos}}$ is large is not the primary target of the proposed protocol. Indeed, the query-free access benchmark performs very well in this case, suggesting that semantic matching is unnecessary to achieve good performance. On the other hand, the proposed semantic query scheme significantly outperforms query-free access for small values of $p_{\text{pos}}$.

Finally, \cref{fig:gaussian_res_query_dim} shows the MD and FA probabilities vs.\ the query dimension using the optimized thresholds. A larger dimension decreases both the probability of MD and of FA. Furthermore, the MD probability for IRSA (dashed line) is generally lower than for ALOHA (solid line), especially at high query dimensions. This suggests that the increased throughput of IRSA makes it possible to take better advantage of the increased discernibility that comes with a higher query dimension. In the case of FA, IRSA is generally better than ALOHA at low query dimensions and similar to ALOHA at high dimensions.

\begin{figure}
  \centering
  \subfloat[MD]{\includetikzimage<results_gaussian_query_dim_vs_md><font=\footnotesize,trim axis right>\label{fig:gaussian_res_query_dim_md}}\hfill
  \subfloat[FA]{\includetikzimage<results_gaussian_query_dim_vs_fa><font=\footnotesize,trim axis left>\label{fig:gaussian_res_query_dim_fa}}
  \caption{Probability of MD \protect\subref{fig:gaussian_res_query_dim_md} and FA \protect\subref{fig:gaussian_res_query_dim_fa} vs. query dimension $l$ under the Gaussian observation model and using optimized thresholds.}\label{fig:gaussian_res_query_dim}
\end{figure}

\begin{figure}
  \centering
  \subfloat[MD, ALOHA]{\includetikzimage<results_modelnet_tau_vs_md><font=\footnotesize,trim axis right>\label{fig:modelnet_res_tau_vs_md_aloha}}\hfill
  \subfloat[MD, IRSA, $\Lambda(x)\!=\!x^3$]{\includetikzimage<results_modelnet_tau_vs_md_irsa><font=\footnotesize,trim axis left>\label{fig:modelnet_res_tau_vs_md_irsa}}\\\vspace{-1em}
  \subfloat[FA, ALOHA]{\includetikzimage<results_modelnet_tau_vs_fa><font=\footnotesize,trim axis right>\label{fig:modelnet_res_tau_vs_fa_aloha}}\hfill
  \subfloat[FA, IRSA, $\Lambda(x)\!=\!x^3$]{\includetikzimage<results_modelnet_tau_vs_fa_irsa><font=\footnotesize,trim axis left>\label{fig:modelnet_res_tau_vs_fa_irsa}}
  \caption{Probability of MD (\protect\subref{fig:modelnet_res_tau_vs_md_aloha}-\protect\subref{fig:modelnet_res_tau_vs_md_irsa}) and FA (\protect\subref{fig:modelnet_res_tau_vs_fa_aloha}-\protect\subref{fig:modelnet_res_tau_vs_fa_irsa}) vs. matching threshold $\tau$ for ALOHA and IRSA under the CNN classification model with $l=8$ and $l=64$. Circles denote performance at the optimized $\tau$ found using the procedure in \cref{sec:cnn_threshold_opt} by maximizing received TPs as a proxy for inference accuracy.}\label{fig:modelnet_res_tau_vs_md_fa}
\end{figure}

\subsection{CNN Classification}
To study a more realistic scenario, we next consider CNN-based semantic data sourcing presented in \cref{sec:neuralnet}. The input images are drawn from the multi-view ModelNet40 dataset~\cite{wu2015modelnet} following the model outlined in \cref{sec:cnn_sensing_model}.
We use the feature extraction layers of the VGG11 model~\cite{simonyan2015vgg11} as the observation feature encoder $f_{\mathrm{enc}}(\cdot)$, which results in $d=25088$ features.
The query and key encoders $f_{\mathrm{q}}(\cdot)$ and $f_{\mathrm{k}}(\cdot)$ are both feed-forward neural networks comprising a single hidden layer with 1024 neurons and ReLU activations. The classifier $f_{\mathrm{ser}}(\cdot)$ is constructed using the classifier layers of the VGG11 model to predict the $|\mathcal{Z}|=40$ categories in the ModelNet40 dataset. Both the VGG11 feature extraction layers and the classifier layers are pretrained on the ImageNet dataset, and then fine-tuned along with the query and key encoder networks on the ModelNet40 dataset with the cross-entropy loss function. Specifically, we split the ModelNet40 training dataset into three disjoint subsets: 80\% of the samples are used for training the CNN, 10\% for validation, and 10\% for calibration (matching threshold optimization). Each of the ModelNet40 subsets are then used to construct samples following the model in \cref{sec:cnn_sensing_model} (samples drawn with replacement). Unless specified, other system parameters are the same as in the GMM experiments. %

\cref{fig:modelnet_res_tau_vs_md_fa} shows the probability of MD and FA vs.\ the matching threshold for query sizes $l=8$ (solid) and $l=64$ (dash-dotted). Similar to the Gaussian case, the MD first decreases with the threshold $\tau$ and then increases, whereas the FA first increases and then decreases. Furthermore, as expected the query dimension $l=64$ generally results in a lower probability of both MD and FA compared to $l=8$, and IRSA performs better than ALOHA. Note that there is a significant gap between the achieved MD/FA and the optimal performance under perfect matching (dashed blue line), suggesting, as in the GMM scenario, that noisy queries and FA significantly impact the performance. However, compared to traditional random access (green dotted line), semantic data sourcing leads to a significant reduction in both MD and FA.

Finally, \cref{fig:modelnet_res_tau_vs_acc} shows the resulting accuracy. As for the GMM, IRSA does not provide significantly higher accuracy compared to ALOHA despite having lower probability of MD and FA, suggesting that bad query examples are the main limitation for accuracy. Overall, the insights obtained from the GMM accurately captures the main behavior of the much more complex CNN, suggesting that the GMM is a good proxy for studying the semantic query-based retrieval problem.

\begin{figure}
  \centering
  \subfloat[ALOHA]{\includetikzimage<results_modelnet_tau_vs_acc><font=\footnotesize,trim axis right>\label{fig:modelnet_res_tau_vs_acc_aloha}}\hfill
  \subfloat[IRSA, $\Lambda(x)=x^3$]{\includetikzimage<results_modelnet_tau_vs_acc_irsa><font=\footnotesize,trim axis left>\label{fig:modelnet_res_tau_vs_acc_irsa}}
  \caption{Inference accuracy vs. matching threshold $\tau$ for \protect\subref{fig:modelnet_res_tau_vs_acc_aloha} ALOHA and \protect\subref{fig:modelnet_res_tau_vs_acc_irsa} IRSA under the CNN classification model. Circles denote performance at the optimized $\tau$ found using the procedure in \cref{sec:cnn_threshold_opt} by maximizing received TPs as a proxy for inference accuracy.}\label{fig:modelnet_res_tau_vs_acc}
\end{figure}

\section{Conclusion}\label{sec:conclusion}
Effective retrieval of sensing data represents a challenge in massive IoT networks. This paper introduces an efficient pull-based data collection protocol for this scenario, termed semantic data sourcing random access. The protocol utilizes semantic queries transmitted by the destination node to enable devices to assess the importance of their sensing data, and relies on modern random access protocols for efficient transmission of the relevant observations. We analyze and explore the tradeoff between MD, FA, and collisions in the random access phase under a tractable GMM, and optimize the matching score threshold to minimize the probability of MD. We also show how the framework can be applied to more realistic but complex scenarios based on a CNN-based model for visual sensing data. Through numerical results, we show that the threshold selection method achieves near-optimal performance in terms of the probability of MD and FA, and that the analysis of the GMM is consistent with results observed from the CNN-based model. Furthermore, the proposed protocol significantly increases the fraction of relevant retrieved observations compared to traditional random access methods, highlighting the effectiveness of semantic query-based data sourcing random access for massive IoT. Future directions include multi-round, feedback-enhanced semantic queries, and energy-efficient semantic queries using, e.g., wake-up radios.

\appendices
\crefalias{section}{appendix}

\section{Proof of \cref{lemma:md_fa}}\label{app:proof_lemma_md_fa}
We first derive the probability of FA, $\varepsilon_{\mathrm{FA},\mathrm{match}}(\tau)$, which occurs if the matching score exceeds $\tau$ when the device observation class $z_m$ is distinct from the query class $z_q$. Let $S_m$ denote the event that device $m$ successfully decodes the query transmission, so that $\Pr(S_m)=1-p_{\mathrm{err},m}^{\mathrm{dl}}$. Since a successful query transmission is required to perform matching, the probability of FA can be expressed as
\begin{align}
  \varepsilon_{\mathrm{FA},\mathrm{match}}(\tau)&=(1-p_{\mathrm{err}}^{\mathrm{dl}})\Pr(\mathrm{FA}|S_m,\tau),\label{eq:lemma1_proof_pfa}
\end{align}
where $\Pr(\mathrm{FA}|S_m,\tau)$ is the probability of FA given successful query reception. This probability is
\begin{align}
  &\Pr(\mathrm{FA}|S_m,\tau)=\Pr(\bar{\chi}(\bm{x}_m,\bm{q})\ge \tau|z_m\neq z_q)\nonumber\\
  &\quad\!=\!\frac{2\sum_{i=1}^{|\mathcal{Z}|}\sum_{j=i+1}^{|\mathcal{Z}|}\Pr(\|\bm{k}_m-\bm{q}\|_2^2\le \tilde{\tau}|z_m=i, z_q=j)}{|\mathcal{Z}|(|\mathcal{Z}|-1)},\label{eq:lemma1_proof_pfa_tmp}
\end{align}
where the last step follows from the uniform distribution of $z_m$ and $z_q$, the monotonicity of $\bar{\chi}(\bm{x}_m,\bm{q})$ in $\|\bm{k}_m-\bm{q}\|_2^2$, and substituting $\tilde{\tau}=d\ln(1/\tau)$.

Given the expressions in \cref{eq:query,eq:key}, the query and key vectors $\bm{q}$ and $\bm{k}_m$ are both independent multivariate Gaussian random variables conditioned on $z_q$ and $z_m$ with mean 
$(\sqrt{d/l})\mathbf{P}_l\boldsymbol{\Sigma}^{-\frac{1}{2}}\boldsymbol{\mu}_{z_q}$ and   $(\sqrt{d/l})\mathbf{P}_l\boldsymbol{\Sigma}^{-\frac{1}{2}}\boldsymbol{\mu}_{z_m}$, respectively, and covariance matrix $(d/l)\mathbf{P}_l\mathbf{P}_l^T=(d/l)\mathbf{I}_l$. Thus $\frac{l}{2d}\|\bm{k}_m-\bm{q}\|_2^2$ follows a noncentral chi-square distribution with $l$ degrees of freedom and noncentrality parameter
\begin{align*}
  &\textstyle\sum_{n=1}^l \left(\left[\left(\sqrt{1/2}\right)\mathbf{P}_l\boldsymbol{\Sigma}^{-\frac{1}{2}}\left(\boldsymbol{\mu}_{z_m}-\boldsymbol{\mu}_{z_q}\right)\right]_n\right)^2\\
  &\qquad=\frac{1}{2}\left(\boldsymbol{\mu}_{z_m}-\boldsymbol{\mu}_{z_q}\right)^T\boldsymbol{\Sigma}^{-\frac{1}{2}}\mathbf{P}_l^T\mathbf{P}_l\boldsymbol{\Sigma}^{-\frac{1}{2}}\left(\boldsymbol{\mu}_{z_m}-\boldsymbol{\mu}_{z_q}\right)\\
  &\qquad\triangleq G_{z_m,z_q}(\mathbf{P}_l)/2.
\end{align*}
Using the expression for the cumulative density function (CDF) of a noncentral chi-square distribution, we obtain
\begin{align}
  &\Pr\left(\|\bm{k}_m-\bm{q}\|_2^2\le \tilde{\tau}l/(2d)\,\middle|\,z_m=i, z_q=j\right)\nonumber\\
  &\qquad=1-Q_{\frac{l}{2}}\left(\sqrt{G_{i,j}(\mathbf{P}_l)/2}, \sqrt{\tilde{\tau}l/(2d)}\right)\nonumber\\
  &\qquad=1-Q_{\frac{l}{2}}\left(\sqrt{G_{i,j}(\mathbf{P}_l)/2}, \sqrt{(l/2)\ln(1/\tau)}\right).\label{eq:cdf_noncentral_chi2}
\end{align}
The final expression for $\varepsilon_{\mathrm{FA},\mathrm{match}}(\tau)$ follows by defining $\tilde{\tau}=l\ln(1/\tau)$ and combining \cref{eq:lemma1_proof_pfa,eq:lemma1_proof_pfa_tmp,eq:cdf_noncentral_chi2}.

The derivation of the probability of MD, $\varepsilon_{\mathrm{MD},\mathrm{match}}(\tau)$, is similar. A MD occurs whenever a device observation from the query class is not transmitted, either due to downlink communication error or an unsuccessful semantic matching. The probability can be expressed as
\begin{align}
  \varepsilon_{\mathrm{MD},\mathrm{match}}(\tau)&=1-\left(1-p_{\mathrm{err}}^{\mathrm{dl}}\right)\left(1-\Pr(\mathrm{MD}|S_m,\tau)\right),\label{eq:lemma1_proof_pmd}
\end{align}
where $\Pr(\mathrm{MD}|S_m,\tau)$ is the probability of MD given successful query reception. This probability is
\begin{align}
  \Pr(\mathrm{MD}|S_m,\tau)&=\Pr(\bar{\chi}(\bm{k}_m,\bm{q})\le \tau|z_m=z_q)\nonumber\\
  &=\Pr(\|\bm{k}_m-\bm{q}\|_2^2\ge \tilde{\tau}|z_m=z_q).\label{eq:lemma1_proof_pmd_tmp}
\end{align}
Following the same argument as before, $\frac{l}{2d}\|\bm{k}_m-\bm{q}\|_2^2$ follows a (central) chi-square distribution with $l$ degrees of freedom. From this we obtain
\begin{align}
  \Pr\left(\|\bm{k}_m-\bm{q}\|_2^2\ge \tilde{\tau}l/(2d)\right)=\frac{\Gamma\left(l/2, (l/4)\ln(1/\tau)\right)}{\Gamma\left(l/2\right)}.\label{eq:cdf_chi2}
\end{align}
The final expression for $\varepsilon_{\mathrm{MD},\mathrm{match}}(\tau)$ follows by combining \cref{eq:lemma1_proof_pmd,eq:lemma1_proof_pmd_tmp,eq:cdf_chi2}.

\section{Proof of \cref{prop:query_selection}}\label{app:proof_prop_query_selection}
We prove the result by showing that problem in \eqref{eq:max_pl} is equivalent to Fisher's LDA~\cite{bishop2006pattern}, which has a well-known solution. We first rewrite the objective function in \cref{eq:max_pl}:
  \begin{align}
    J(\mathbf{P}_l)&=\textstyle\frac{2}{|\mathcal{Z}|(|\mathcal{Z}|-1)}\sum_{i=1}^{|\mathcal{Z}|}\sum_{j=i+1}^{|\mathcal{Z}|}G_{i,j}(\mathbf{P}_l)\nonumber\\
    &=\textstyle\frac{1}{2}\sum_{i=1}^{|\mathcal{Z}|}\sum_{j=1}^{|\mathcal{Z}|}\|\mathbf{P}_l\boldsymbol{\Sigma}^{-\frac{1}{2}}\left(\boldsymbol{\mu}_i-\boldsymbol{\mu}_j\right)\|_2^2\nonumber\\
    &=\textstyle\frac{1}{2}\sum_{i=1}^{|\mathcal{Z}|}\sum_{j=1}^{|\mathcal{Z}|}\|\mathbf{P}_l\left(\tilde{\boldsymbol{\mu}}_i-\tilde{\boldsymbol{\mu}}_j\right)\|_2^2,\label{eq:disc_gain_prop_eq1}
  \end{align}
  where $\tilde{\boldsymbol{\mu}}_i=\boldsymbol{\Sigma}^{-\frac{1}{2}}(\boldsymbol{\mu}_i-\bar{\boldsymbol{\mu}})$ and  $\bar{\boldsymbol{\mu}}=\frac{1}{|\mathcal{Z}|}\sum_{i=1}^{|\mathcal{Z}|}\boldsymbol{\mu}_i$. \Cref{eq:disc_gain_prop_eq1} can in turn be written
  \begin{align*}
     J(\mathbf{P}_l)&=\!\textstyle\frac{1}{2}\sum_{i=1}^{|\mathcal{Z}|}\sum_{j=1}^{|\mathcal{Z}|}\Tr\left(\mathbf{P}_l\left(\tilde{\boldsymbol{\mu}}_i-\tilde{\boldsymbol{\mu}}_j\right)\left(\tilde{\boldsymbol{\mu}}_i-\tilde{\boldsymbol{\mu}}_j\right)^T\mathbf{P}_l^T\right)\nonumber\\
    &=\!\textstyle\frac{1}{2}\!\Tr\!\left(\mathbf{P}_l\!\left(\sum_{i=1}^{|\mathcal{Z}|}\sum_{j=1}^{|\mathcal{Z}|}\left(\tilde{\boldsymbol{\mu}}_i\!-\!\tilde{\boldsymbol{\mu}}_j\right)\left(\tilde{\boldsymbol{\mu}}_i\!-\!\tilde{\boldsymbol{\mu}}_j\right)^T\right)\mathbf{P}_l^T\right).
\end{align*}
Next, we note that
  \begin{align*}
    &\left[\textstyle\sum_{i=1}^{|\mathcal{Z}|}\sum_{j=1}^{|\mathcal{Z}|}\left(\tilde{\boldsymbol{\mu}}_i-\tilde{\boldsymbol{\mu}}_j\right)\left(\tilde{\boldsymbol{\mu}}_i-\tilde{\boldsymbol{\mu}}_j\right)^T\right]_{k,l}\nonumber\\
    &\qquad=\textstyle\sum_{i=1}^{|\mathcal{Z}|}\sum_{j=1}^{|\mathcal{Z}|}\left(\tilde{\boldsymbol{\mu}}_i[k]-\tilde{\boldsymbol{\mu}}_j[k]\right)\left(\tilde{\boldsymbol{\mu}}_i[l]-\tilde{\boldsymbol{\mu}}_j[l]\right)^T\nonumber\\
    &\qquad=\textstyle 2|\mathcal{Z}|\sum_{i=1}^{|\mathcal{Z}|}\left(\tilde{\boldsymbol{\mu}}_i[k]\tilde{\boldsymbol{\mu}}_i[l]-\tilde{\boldsymbol{\mu}}_i[l]\sum_{j=1}^{|\mathcal{Z}|}\tilde{\boldsymbol{\mu}}_j[l]\right)\nonumber\\
    &\qquad=\textstyle 2|\mathcal{Z}|\sum_{i=1}^{|\mathcal{Z}|}\tilde{\boldsymbol{\mu}}_i[k]\tilde{\boldsymbol{\mu}}_i[l],
  \end{align*}
  where the last step follows from the fact that $\sum_{j=1}^{|\mathcal{Z}|}\tilde{\boldsymbol{\mu}}_j=\sum_{j=1}^{|\mathcal{Z}|}\boldsymbol{\Sigma}^{-\frac{1}{2}}(\boldsymbol{\mu}_i-\bar{\boldsymbol{\mu}})=\boldsymbol{0}$. It follows that
  \begin{align*}
    \textstyle\sum_{i=1}^{|\mathcal{Z}|}\sum_{j=1}^{|\mathcal{Z}|}\left(\tilde{\boldsymbol{\mu}}_i-\tilde{\boldsymbol{\mu}}_j\right)\left(\tilde{\boldsymbol{\mu}}_i-\tilde{\boldsymbol{\mu}}_j\right)^T = 2|\mathcal{Z}|\sum_{i=1}^{|\mathcal{Z}|}\tilde{\boldsymbol{\mu}}_i\tilde{\boldsymbol{\mu}}_i^T,
 \end{align*}
 and thus
 \begin{align*}
   J(\mathbf{P}_l)&=\textstyle|\mathcal{Z}|\Tr\left(\mathbf{P}_l\left(\sum_{i=1}^{|\mathcal{Z}|}\tilde{\boldsymbol{\mu}}_i\tilde{\boldsymbol{\mu}}_i^T\right)\mathbf{P}_l^T\right)\nonumber\\
   &=|\mathcal{Z}|\Tr\left(\mathbf{P}_l\mathbf{W}\mathbf{P}_l^T\right),
 \end{align*}
 where $\mathbf{W}=\sum_{i=1}^{|\mathcal{Z}|}\tilde{\boldsymbol{\mu}}_i\tilde{\boldsymbol{\mu}}_i^T=\sum_{i=1}^{|\mathcal{Z}|}\boldsymbol{\Sigma}^{-\frac{1}{2}}(\boldsymbol{\mu}_i-\bar{\boldsymbol{\mu}_i})(\boldsymbol{\mu}_i-\bar{\boldsymbol{\mu}_i})^T\boldsymbol{\Sigma}^{-\frac{1}{2}}$. Disregarding the constant $|\mathcal{Z}|$, maximizing $\Tr\left(\mathbf{P}_l\mathbf{W}\mathbf{P}_l^T\right)$ with the constraint $\mathbf{P}_l\mathbf{P}_l^T=\mathbf{I}_l$ is equivalent to Fisher's linear discriminant analysis with covariance matrix $\mathbf{I}_d$. This problem has the well-known solution given by
 \begin{equation*}
   \mathbf{P}_l = \begin{bmatrix}
     \bm{v}_1^T & \bm{v}_2^T & \cdots & \bm{v}_l^T
   \end{bmatrix}^T,
 \end{equation*}
where $\bm{v}_1,\bm{v}_2,\ldots,\bm{v}_d\in\mathbb{R}^{d}$ are the eigenvectors of $\mathbf{W}$ ordered by the magnitude of their corresponding eigenvalues~\cite{bishop2006pattern}.

\section{Proof of \cref{prop:opt_tau}}\label{app:proof_prop_opt_tau}
We first prove that the expression in \cref{eq:opt_tau} is a necessary and sufficient condition for optimality, and then show that there exists a solution for $0<\tau\le 1$.

The derivative of \cref{eq:approx_ntp} is
\begin{align*}
  &\frac{\df}{\df\tau}\left(\bar{\lambda}_{\mathrm{TP}}(\tau)e^{-\bar{\lambda}(\tau)/L_{\mathrm{ul}}}\right)\nonumber\\
  &\quad=e^{-\bar{\lambda}(\tau)/L_{\mathrm{ul}}}{\bar{\lambda}_{\mathrm{TP}}}'(\tau)+\bar{\lambda}_{\mathrm{TP}}(\tau)\left(\frac{\df}{\df\tau}e^{-\bar{\lambda}(\tau)/L_{\mathrm{ul}}}\right)\nonumber\\
  &\quad=e^{-\bar{\lambda}(\tau)/L_{\mathrm{ul}}}{\bar{\lambda}_{\mathrm{TP}}}'(\tau)-\frac{\bar{\lambda}_{\mathrm{TP}}(\tau)e^{-\bar{\lambda}(\tau)/L_{\mathrm{ul}}}}{L_{\mathrm{ul}}}\bar{\lambda}'(\tau)\nonumber\\
    &\quad=\!\frac{e^{-\bar{\lambda}(\tau)/L_{\mathrm{ul}}}}{L_{\mathrm{ul}}}\!\left(\!\left(L_{\mathrm{ul}}\!-\!\bar{\lambda}_{\mathrm{TP}}(\tau)\right){\bar{\lambda}_{\mathrm{TP}}}'(\tau)\!-\!\bar{\lambda}_{\mathrm{TP}}(\tau){\bar{\lambda}_{\mathrm{FA}}}'(\tau)\right).
\end{align*}
By setting the expression equal to zero and noting that $e^{-\bar{\lambda}(\tau)/L_{\mathrm{ul}}}>0$ for $\bar{\lambda}(\tau)<\infty$, we obtain the condition
\begin{align*}
 \psi(\tau)\overset{\triangle}{=} L_{\mathrm{ul}}-\bar{\lambda}_{\mathrm{TP}}(\tau)\left(1+\frac{{\bar{\lambda}_{\mathrm{FA}}}'(\tau)}{{\bar{\lambda}_{\mathrm{TP}}}'(\tau)}\right)=0.
\end{align*}
To simplify the subsequent notation, let $g(\tau)=l\ln(1/\tau)$. Using the definition in \cref{eq:bar_lambda_td}, we have
\begin{align}
  {\bar{\lambda}_{\mathrm{TP}}}'(\tau)&=-\frac{Mp_{\mathrm{pos}}(1-p_{\mathrm{err}}^{\mathrm{dl}})lg(\tau)^{l/2-1}}{ 2^{l}\Gamma(l/2)\tau^{1-l/4}},\label{eq:df_lambda_td}
\end{align}
where we applied the chain rule and used that $1-\Gamma(l/2, \tilde{\tau}/4)/\Gamma(l/2)$ is the CDF of a chi-squared distribution, whose derivative is the PDF. Using a similar approach with the definition in \cref{eq:bar_lambda_fa}, but using the noncentral chi-squared distribution, we find
\begin{align}
  {\bar{\lambda}_{\mathrm{FA}}}'(\tau)&\!=\!-\!\frac{M(1-p_{\mathrm{pos}})(1-p_{\mathrm{err}}^{\mathrm{dl}})lg(\tau)^{l/2-1}}{2|\mathcal{Z}|(|\mathcal{Z}|-1)\tau^{1-l/4}}\phi\left(g(\tau)\right)\label{eq:df_lambda_fa}
\end{align}
where
\begin{multline*}
  \phi\left(g(\tau)\right)=g(\tau)^{1/2-l/4}\\
  \textstyle\times{\sum_{i=1}^{|\mathcal{Z}|}\sum_{j=i+1}^{|\mathcal{Z}|}}\left[\frac{e^{-\frac{1}{4}G_{i,j}}}{G_{i,j}^{l/4-1/2}}
I_{l/2-1}\left(\frac{1}{2}\sqrt{G_{i,j}g(\tau)}\right)\right],
\end{multline*}
$I_{n}(x)$ is the modified Bessel function of the first kind, and we used the shorthand $G_{i,j}$ instead of $G_{i,j}(\mathbf{P}_l)$ for convenience.
Inserting \cref{eq:df_lambda_td,eq:df_lambda_fa} into the expression for $\psi(\tau)$ we obtain
\begin{align*}
  \psi(\tau)&=L_{\mathrm{ul}}-\bar{\lambda}_{\mathrm{TP}}(\tau)\left(1+\frac{(1-p_{\mathrm{pos}})2^{l-1}\Gamma(l/2)}{p_{\mathrm{pos}}|\mathcal{Z}|(|\mathcal{Z}|-1)}\phi\left(g(\tau)\right)\right)\\
  &=L_{\mathrm{ul}}-M(1-p_{\mathrm{err}}^{\mathrm{dl}})\gamma\left(\frac{l}{2},\frac{g(\tau)}{4}\right)\nonumber\\
  &\qquad\times\left(p_{\mathrm{pos}}+\frac{(1-p_{\mathrm{pos}})2^{l-1}\Gamma(l/2)}{|\mathcal{Z}|(|\mathcal{Z}|-1)}\phi\left(g(\tau)\right)\right),
\end{align*}
where the last step follows from \cref{eq:bar_lambda_td,eq:md} and the identity $\Gamma(s,z)/\Gamma(s)=1-\gamma(s,z)$.

We next to show that $\psi(\tau)$ is strictly increasing in $\tau$. Since the function is continuous in the interval $0<\tau\le 1$, strict monotonicity implies that any root of $\psi(\tau)$ (if any) is unique, and thus maximizes \cref{eq:approx_ntp}. We first note that to be strictly increasing in $\tau$ in the interval $0<\tau\le 1$, the function must be strictly decreasing in $g(\tau)$. To this end, since all coefficients are positive it suffices to show that $\gamma(l/2,g(\tau)/4)$ and $\phi(g(\tau))$ are both strictly increasing in $g(\tau)$. Clearly, this holds for $\gamma(l/2,g(\tau)/4)$. For $\phi(g(\tau))$, it is sufficient to show that
\begin{equation*}
  g(\tau)^{1/2-l/4}I_{l/2-1}\left(\frac{1}{2}\sqrt{G_{i,j}g(\tau)}\right)
\end{equation*}
is increasing in $g(\tau)$ for all $i,j$. Applying the identity $I_{n}(x)=\left(\frac{x}{2}\right)^n\sum_{k=0}^{\infty}\frac{(x^2/4)^k}{k!\Gamma(n+k+1)}$~\cite[eq. 9.6.10]{abramowitz1970handbook}, we obtain
\begin{multline*}
  g(\tau)^{1/2-l/4}I_{l/2-1}\left(\frac{1}{2}\sqrt{G_{i,j}g(\tau)}\right)\\
  \textstyle=
2^{2-l}g(\tau)^{3(l-1)/4}G_{i,j}^{l/4-1/2}\sum_{k=0}^{\infty}\frac{(G_{i,j}g(\tau)/16)^k}{k!\Gamma(l/2+k)}.
\end{multline*}
Since this function is strictly increasing in $g(\tau)$ when $l\ge 1$ and $G_{i,j}>0$ and zero when $G_{i,j}=0$, we conclude that $\phi(g(\tau))$ is strictly increasing in $g(\tau)$, and thus $\psi(\tau)$ is strictly increasing in $\tau$ for $l\ge 1$ and $\sum_{i=1}^{|\mathcal{Z}|}\sum_{j=i+1}^{|\mathcal{Z}|}G_{i,j}>0$.

Finally, to see that $\psi(\tau)$ has a root in the interval $0<\tau\le 1$, we note that under the conditions stated above,
\begin{equation*}
  \begin{array}{ll}
    \psi(\tau)\to-\infty, & \textrm{as }\tau\to 0,\\
    \psi(\tau)\to L_{\mathrm{ul}}, &\textrm{as }\tau\to 1.
  \end{array}
\end{equation*}
By the intermediate value theorem, $\psi(\tau)$ must have a root in $0<\tau<1$. Furthermore, since $\psi(\tau)$ is strictly increasing, this root is unique. The proof is completed by defining $\tilde{\tau}=g(\tau)$.

\input{main.bbl}

\begin{IEEEbiography}[{\includegraphics[width=1in,height=1.25in,clip,keepaspectratio]{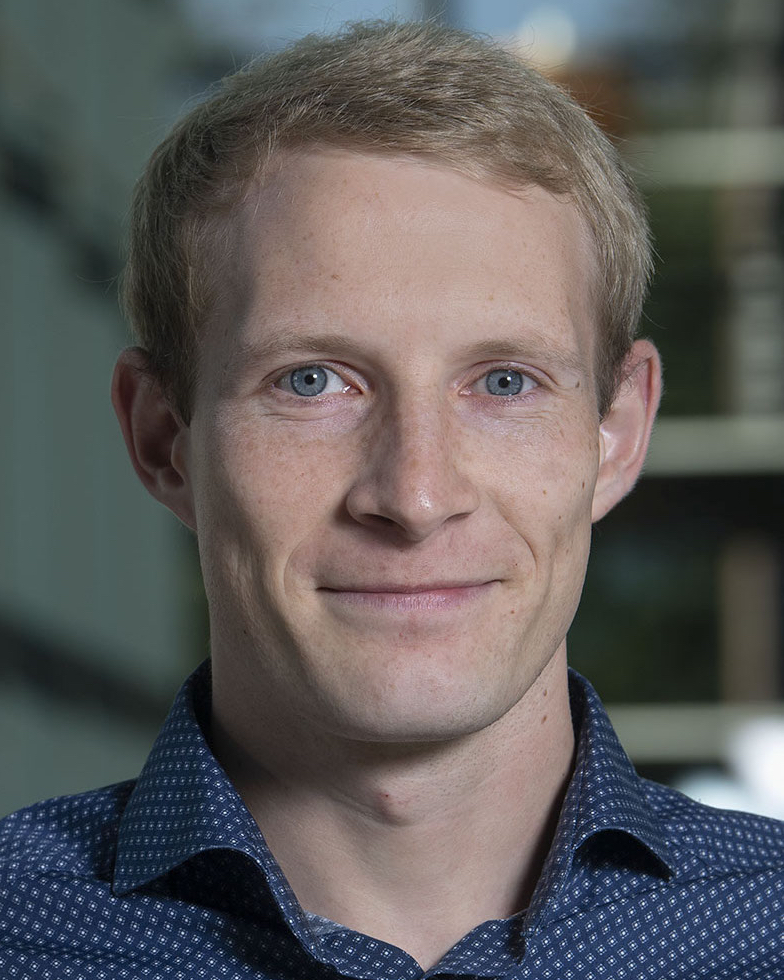}}]{Anders E. Kal{\o}r} (Member, IEEE) received the B.Sc. and M.Sc. degrees in computer engineering and the Ph.D. degree in wireless communications from Aalborg University, Denmark, in 2015, 2017, and 2022, respectively. He is currently a Project Assistant Professor at the Department of Information and Computer Science, Keio University, Japan. He was a postdoctoral researcher at The University of Hong Kong from 2022 to 2023, and at Aalborg University from 2023 to 2024, supported by an International Postdoc grant from the Independent Research Fund Denmark (IRFD). He was awarded the Spar Nord Foundation Research Award for his Ph.D. project in 2023. His current research interests include communication theory and the intersection between wireless communications, machine learning, and information theory for IoT.
\end{IEEEbiography}

\begin{IEEEbiography}[{\includegraphics[width=1in,height=1.25in,clip,keepaspectratio]{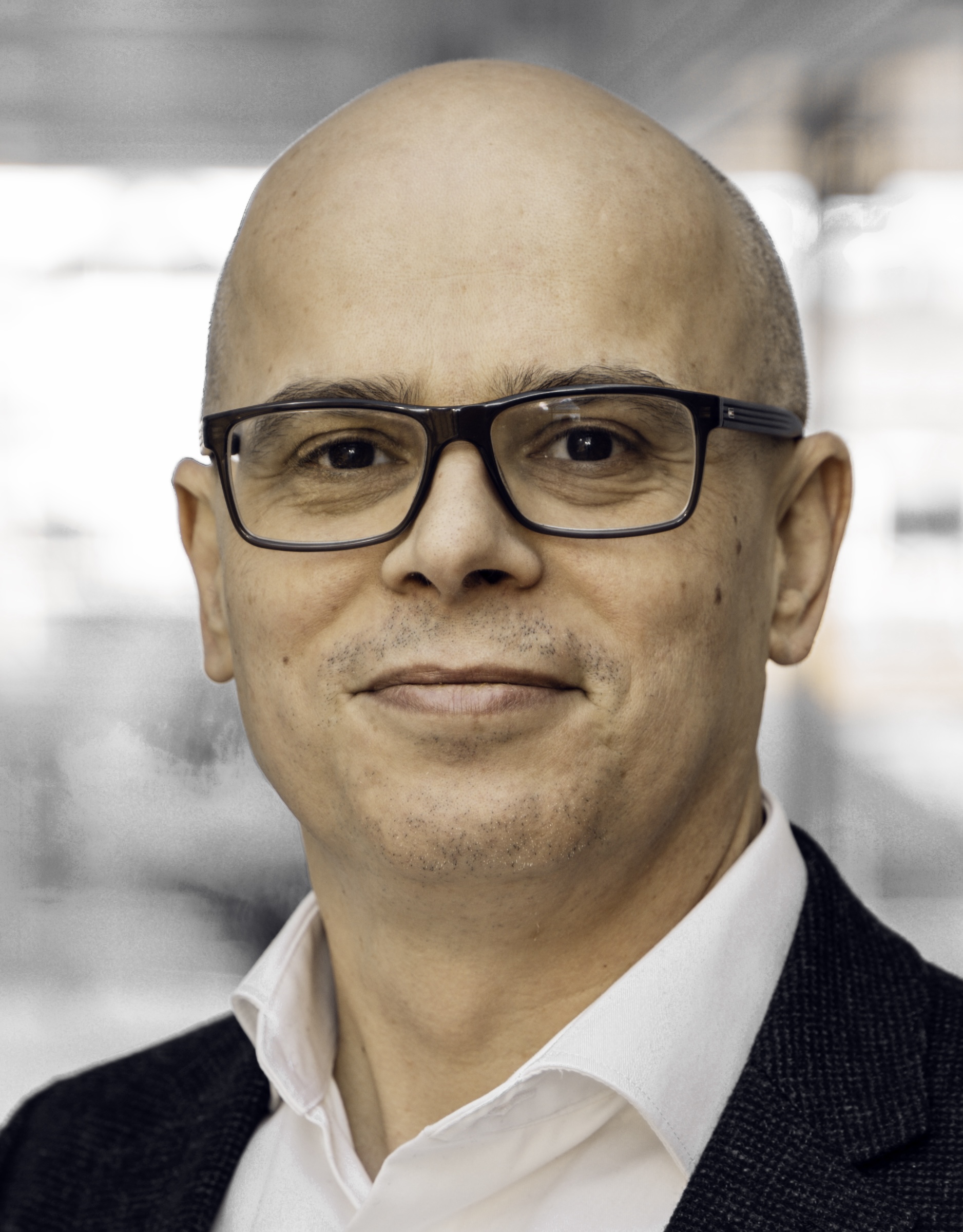}}]{Petar Popovski} (Fellow, IEEE) is a Professor at Aalborg University, where he is the director of the Center of Excellence CLASSIQUE (Classical Communication in the Quantum Era). He holds a position of a Visiting Excellence Chair at the University of Bremen and a Visiting Professor at the University of Sts. Cyril and Methodius (UKIM) in Skopje. He received his Dipl.-Ing (1997) and M. Sc. (2000) degrees in communication engineering from UKIM and the Ph.D. degree (2005) from Aalborg University. He is a Fellow of the IEEE. He received technical recognition/achievement award from multiple Technical Committees of the IEEE Communications Society: Smart Grid Communications (2019), Wireless Communications (2024), and Satellite and Space Communications (2025). In addition, he received ERC Consolidator Grant (2015), the Danish Elite Researcher award (2016), IEEE Fred W. Ellersick prize (2016), IEEE Stephen O. Rice prize (2018), the Danish Telecommunication Prize (2020), and Villum Investigator Grant (2021). He was a Member at Large at the Board of Governors in IEEE Communication Society and Chair of the IEEE Communication Theory Technical Committee. His research interests are in the area of wireless communication and communication theory. He authored the book ``Wireless Connectivity: An Intuitive and Fundamental Guide'', published by Wiley in 2020. He is currently an Editor-in-Chief of IEEE JOURNAL ON SELECTED AREAS IN COMMUNICATIONS.
\end{IEEEbiography}

\begin{IEEEbiography}[{\includegraphics[width=1in,height=1.25in,clip,keepaspectratio]{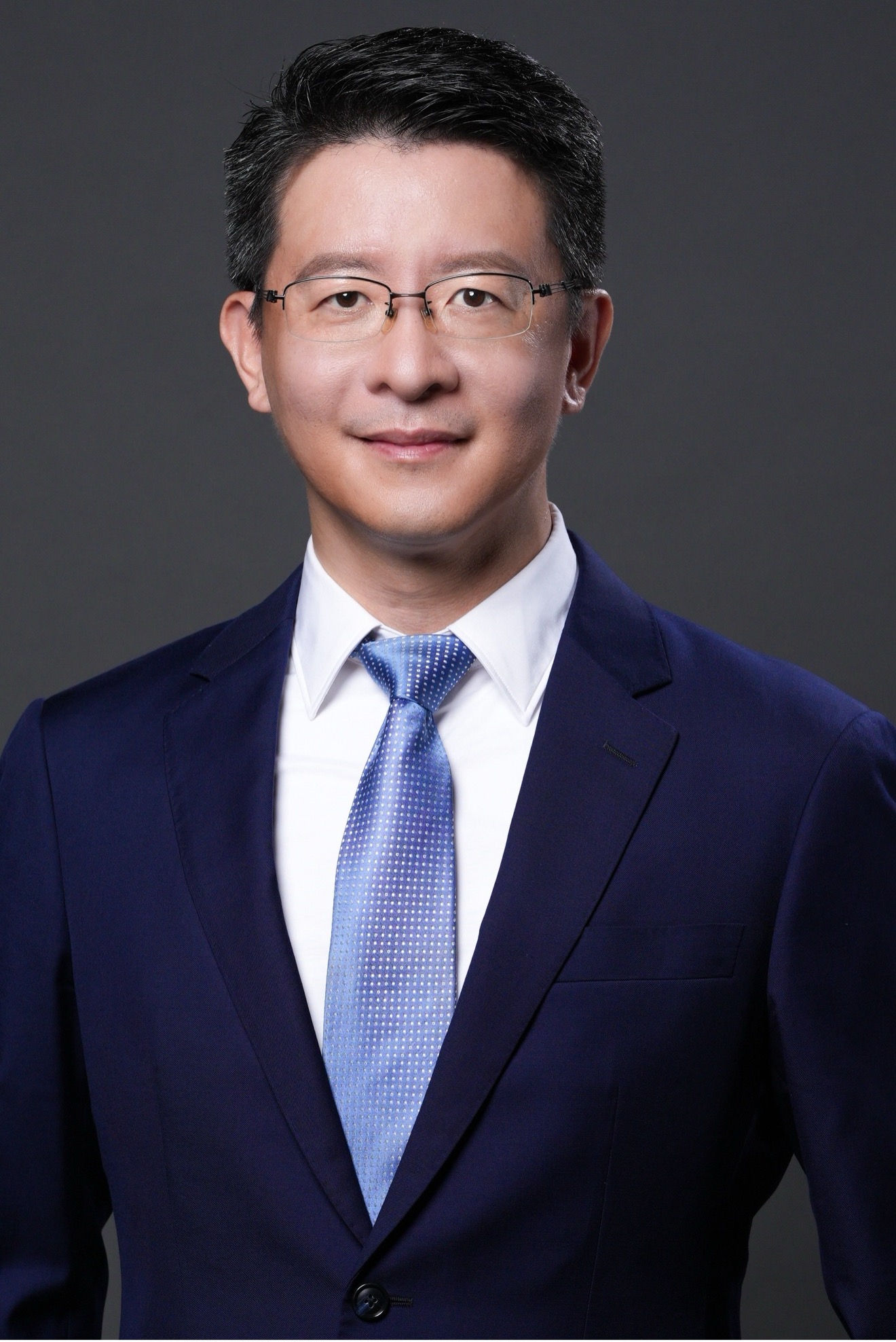}}]{Kaibin Huang} (Fellow, IEEE) received the B.Eng. and M.Eng. degrees from the National University of Singapore and the Ph.D. degree from The University of Texas at Austin, all in electrical engineering. He is the Philip K H Wong Wilson K L Wong Professor in Electrical Engineering and the Department Head at the Dept. of Electrical and Electronic Engineering, The University of Hong Kong (HKU), Hong Kong. His work was recognized with seven Best Paper awards from the IEEE Communication Society. He is a member of the Engineering Panel of Hong Kong Research Grants Council (RGC) and a RGC Research Fellow (2021 Class). He has served on the editorial boards of five major journals in the area of wireless communications and co-edited ten journal special issues. He has been active in organizing international conferences such as the 2014, 2017, and 2023 editions of IEEE Globecom, a flagship conference in communication. He has been named as a Highly Cited Researcher by Clarivate in the last six years (2019-2024) and an AI 2000 Most Influential Scholar (Top 30 in Internet of Things) in 2023-2024. He was an IEEE Distinguished Lecturer. He is a Fellow of the U.S. National Academy of Inventors.
\end{IEEEbiography}

\end{document}